\documentclass[reprint,superscriptaddress,showpacs,amsmath,amssymb,aps,pra]{revtex4-1}
\usepackage{graphicx}
\usepackage{mathrsfs}
\usepackage{subfigure}
\usepackage{amsthm}
\usepackage{hyperref}
\hypersetup{colorlinks=true, citecolor=blue, urlcolor=blue, linkcolor=blue}
\begin{document}
\title{Assisted quantum simulation of open quantum systems}
\author{Jin-Min Liang}
\email{jmliang@cnu.edu.cn}
\affiliation{School of Mathematical Sciences, Capital Normal University, Beijing 100048, China}
\author{Qiao-Qiao Lv}
\email{lvqq@cnu.edu.cn}
\affiliation{School of Mathematical Sciences, Capital Normal University, Beijing 100048, China}
\author{Zhi-Xi Wang}
\email{wangzhx@cnu.edu.cn}
\affiliation{School of Mathematical Sciences, Capital Normal University, Beijing 100048, China}
\author{Shao-Ming Fei}
\email{feishm@cnu.edu.cn}
\affiliation{School of Mathematical Sciences, Capital Normal University, Beijing 100048, China}

\begin{abstract}
Universal quantum algorithms (UQA) implemented on fault-tolerant quantum computers are expected to achieve an exponential speedup over classical counterparts. However, the deep quantum circuits make the UQA implausible in the current era. With only the noisy intermediate-scale quantum (NISQ) devices in hand, we introduce the quantum-assisted quantum algorithm, which reduces the circuit depth of UQA via NISQ technology. Based on this framework, we present two quantum-assisted quantum algorithms for simulating open quantum systems, which utilize two parameterized quantum circuits to achieve a short-time evolution. We propose a variational quantum state preparation method, as a subroutine to prepare the ancillary state, for loading a classical vector into a quantum state with a shallow quantum circuit and logarithmic number of qubits. We demonstrate numerically our approaches for a two-level system with an amplitude damping channel and an open version of the dissipative transverse field Ising model on two sites.
\medskip

\textbf{Keywords:} Quantum simulation, Open quantum system, Variational quantum algorithm
\end{abstract}

\maketitle
\tableofcontents
\section{Introduction}
The novel advantages of fault-tolerant quantum (FTQ) computers, a hardware platform to perform universal quantum algorithms (UQA), have been theoretically and experimentally demonstrated in simulation of many-body quantum systems \cite{Benioff1980The,Feynman2018Simulating,Hempel2018Quantum,Joo2022Commutation}, prime factorization \cite{Shor1997Polynomial}, linear equation solving \cite{Harrow2009Quantum} and machine learning problems \cite{Rebentrost2014Quantum,Duan2022Hamiltonian}. However, enough long coherent time induced by the deep quantum circuits could be a catastrophic obstacles to the execution of UQA \cite{Unruh1995Maintaining,Peruzzo2014Variational}. Although quantum error correction (QEC) is a candidate to mitigate this fatal problem \cite{Devitt2013Quantum}, performing QEC requires the manipulation on many additional qubits and quantum gates \cite{Lidar2013Quantum}. Besides the aforementioned issues, when we implement an UQA in tackling quantum machine learning tasks, the quantum state encoding of classical data and the readout of quantum state are additional crucial challenges in benefiting the quantum advantages \cite{Biamonte2017Quantum}.

In the near term, the noisy intermediate-scale (50-100 qubit) quantum (NISQ) computers applying variational quantum algorithms (VQAs) avoid the implementation of QEC and have shallow quantum circuit compared with the FTQ computers \cite{Preskill2018Quantum}. A variety of high-impact applications of NISQ devices have been studied in many-body quantum system simulations \cite{Lierta2021Meta,Cerezo2021Variational,Gibbs2021Long,Bharti2022Noisy,Lau2022NISQ} and machine learning \cite{Shingu2021Boltzmann,Wei2022A}. However, there have been no corresponding VQAs for some special problems such as the direct estimation of energy difference between two structures in chemistry, although a promising UQA has been already developed \cite{Sugisaki2021A}. In this scenario, realizing UQA with NISQ hardware is an urgent task. Unfortunately, direct implementation of UQA on NISQ devices is almost impossible due to deep quantum circuit \cite{Leymann2020The}. Moreover, UQA and VQAs have different advantages. Especially for quantum simulation, variational quantum simulation has shallow circuit depth but requires to learn different parameterized quantum circuits (PQCs) for different initial states \cite{Bharti2022Noisy}. The universal Trotterization approach simulates the time dynamic of arbitrary initial state with a fixed unitary but has deep circuit depth.

In this work, we introduce an algorithm framework, the \textit{quantum-assisted quantum algorithm}, which follows the structure of the corresponding UQA and employs the NISQ technology to reduce the circuit depth of the UQA. As an important subroutine, we propose a variational quantum state preparation (VQSP) for loading an arbitrary classical data into an amplitude encoding state via learning a PQC. As an application of the proposed algorithm framework, we introduce two quantum-assisted quantum algorithms for simulating open quantum systems (OQS). The study of OQS allows us to understand a rich variety of phenomena including nonequilibrium phase transitions \cite{Pizorn2013One}, biological systems \cite{Marais2013Decoherence}, and the nature of dissipation and decoherence \cite{Daley2014Quantum}. Compared with prior variational quantum simulations \cite{Endo2020Variational,Haug2020Generalized} and Trotterization approach, our simulation approach can evolve an arbitrary initial state by learning two PQCs and reduce the circuit depth.

\section{Results}
\subsection{Quantum-assisted quantum algorithm}\label{subsecI}
For a given problem the UQA performed on an $n$-qubit system is formulated in terms of (i) an initial state $|\Phi_{\textrm{in}}\rangle=\mathcal{U}_{\textrm{in}}|0^{\otimes n}\rangle$, (ii) an evolution unitary $\mathcal{U}_{\textrm{e}}$, (iii) a measurement protocol $\mathcal{P}$ for the output state $|\Phi_{\textrm{out}}\rangle=\mathcal{U}_{\textrm{e}}|\Phi_{\textrm{in}}\rangle$. A quantum-assisted quantum algorithm, as shown in Fig. 1, seeks to solve the problem by learning two PQCs $\widetilde{\mathcal{U}}_{\textrm{in}}(\boldsymbol{\alpha}_{\textrm{opt}})$ and $\widetilde{\mathcal{U}}_{\textrm{e}}(\boldsymbol{\beta}_{\textrm{opt}})$ determined by parameters $\boldsymbol{\alpha}_{\textrm{opt}}$ and $\boldsymbol{\beta}_{\textrm{opt}}$ for $\mathcal{U}_{\textrm{in}}$ and $\mathcal{U}_{\textrm{e}}$ such that
\begin{align}
|\langle\Phi_{\textrm{in}}|\widetilde{\mathcal{U}}_{\textrm{in}}(\boldsymbol{\alpha}_{\textrm{opt}})
|0^{\otimes n}\rangle|^{2}&\geq1-\epsilon_{\textrm{in}},\label{Compiling1}\\
\int_{\psi}|\langle\psi|\mathcal{U}_{\textrm{e}}^{\dag}
\widetilde{\mathcal{U}}_{\textrm{e}}(\boldsymbol{\beta}_{\textrm{opt}})|\psi\rangle|^{2}d\psi&\geq1-\epsilon_{\textrm{e}},
\label{Compiling2}
\end{align}
where the errors $0<\epsilon_{\textrm{in}},\epsilon_{\textrm{e}}\ll1$. Eq. (\ref{Compiling2}) is the fidelity averaged over the Haar distribution \cite{Khatri2019Quantum}.
\begin{figure}[ht]
\centering
\includegraphics[scale=0.5]{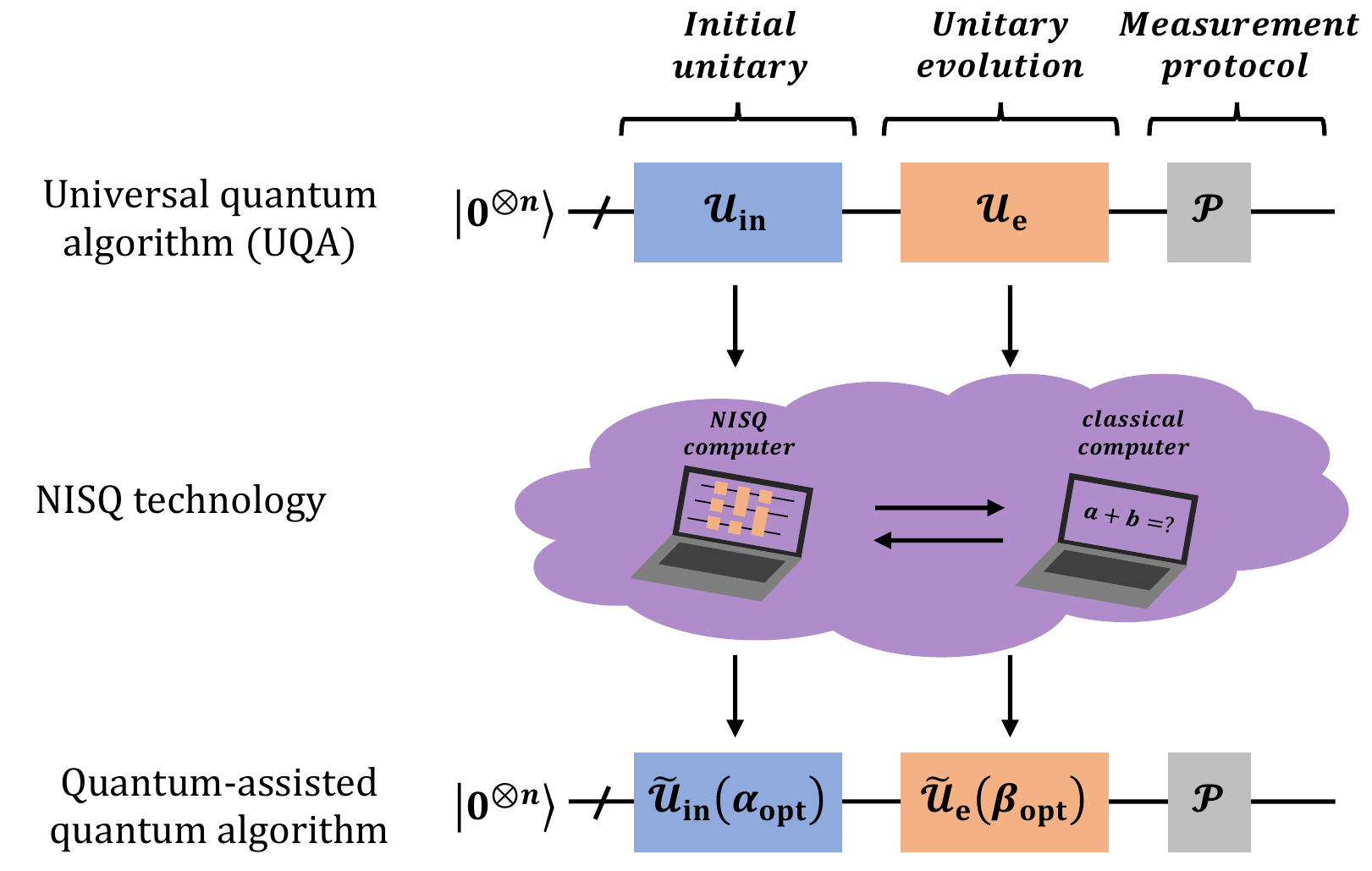}
\caption{Quantum-assisted quantum algorithm.}
\label{NISQ-assisted}
\end{figure}

The quantum-assisted quantum algorithm tailors the parts of UQA and renders it suitable on the NISQ era. There are two crucial problems: (a) How to prepare a specific state with NISQ technology. For almost all VQAs, the initial state is often some easy-prepared one, i.e., a product state with all qubits in the $|0\rangle$ state. Different initial states may not affect the final solution, although a good initial state allows for the VQA to start the search in a region of the parameter space that is closer to the optimum \cite{Bharti2022Noisy}. However, in UQA the initial state may be a specific state determined by the specific problem. For instance, in the Harrow-Hassidim-Lloyd (HHL) algorithm \cite{Harrow2009Quantum}, the initial state is a right-side vector state $|b\rangle$. A similar problem is also encountered in the realm of quantum machine learning \cite{Rebentrost2014Quantum}. It has been demonstrated that an exact universally algorithm would need at least $\mathcal{O}(n)$ qubits and $\mathcal{O}(2^{n})$ operators to prepare a general $2^{n}$-dimensional classical vector \cite{Plesch2011Quantum}. The circuit depth is exponential in the number of qubits and thus such state preparation approach is unsuitable for NISQ devices. In next subsection, we achieve the state preparation via introducing a hybrid quantum-classical approach. (b) How to compile the target unitary $\mathcal{U}_{e}$ to a unitary $\widetilde{\mathcal{U}}_{\textrm{e}}(\boldsymbol{\beta}_{\textrm{opt}})$ with short-depth gate sequences. Solving the problem (b) is the goal of quantum compiling \cite{Chong2017Programming,Haner2018A,Madden2021Best,Mizuta2022Local}. There have spectacular advancements in the field of learning a (possibly unknown) unitary with a lower-depth unitary including VQAs \cite{Sharma2020Noise,Xu2021Variational} and machine learning methods \cite{Zhang2020Topological,He2021Variational}.

\subsection{Variational quantum state preparation}\label{subsecII}
Given a normalized vector $\vec{x}=(x_0,x_1,\cdots,x_{D-1})\in\mathbb{C}^{D}$, $\sum_{j=0}^{D-1}x_{j}^2=1$, and $D=2^{d}$, $d$ is an integer. A quantum state preparation (QSP) aims at preparing a $d$-qubit quantum state
\begin{align}
|x\rangle=\sum_{j=0}^{D-1}x_{j}|j\rangle=U_{x}|0^{\otimes d}\rangle,
\end{align}
by acting a unitary $U_{x}$ on a tensor product state $|0^{\otimes d}\rangle$. We here consider three different cases.

\emph{Case 1: all nonnegative or all negative vector.} We first consider a normalized vector $\vec{x}=(x_0,\cdots,x_j,\cdots,x_{D-1})$, $0\leq x_j\in\mathbb{R}$. All negative vector is covered by adding a global phase $-1$. Algorithm 1, first presented in \cite{Liang2022Improved}, shows the detailed process of a hybrid quantum-classical algorithm to prepare state $|x\rangle$.

\emph{Algorithm 1:} Variational quantum state preparation (VQSP).

\textbf{Input:} an initial state $|0^{\otimes d}\rangle$ and a PQC $U(\boldsymbol\theta)$.

(1) Measure the parameterized quantum state $|\Phi(\boldsymbol\theta)\rangle=U(\boldsymbol\theta)|0^{\otimes d}\rangle$ in the standard basis $\{|j\rangle\}$ and obtain the probability of seeing result $j$, $P_{j}^{\boldsymbol\theta}$.

(2) Estimate the cost function $F_1(\boldsymbol\theta)$ (Eq. \ref{CostFun1}).

(3) Find the minimal value of $F_1(\boldsymbol\theta)$ by classical optimization algorithms.

\textbf{Output:} an optimal parameter $\boldsymbol\theta_{\textrm{opt}}$ and the amplitude encoded state $|x\rangle\approx|\Phi(\boldsymbol\theta_{\textrm{opt}})\rangle=U(\boldsymbol\theta_{\textrm{opt}})|0^{\otimes d}\rangle$.

\begin{figure}[ht]
\includegraphics[scale=0.3]{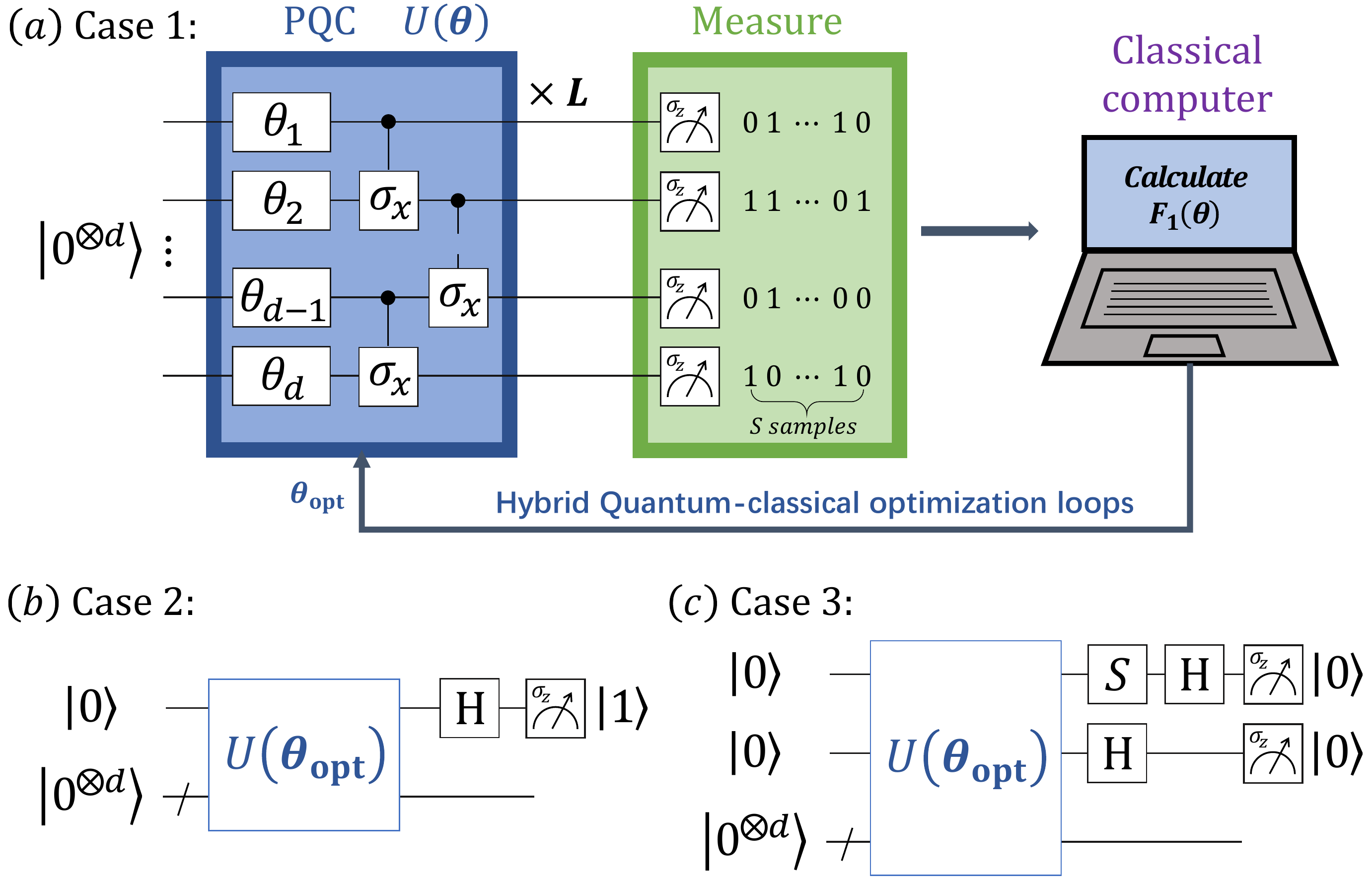}
\caption{Overview of the amplitude encoding in three cases. The PQC consists of single-qubit quantum gate $R_y(\theta_k)=e^{-\iota\theta_k/2\sigma_y}$, $\iota^2=-1$, and two-qubit entangled gates (\textrm{CNOT}).}
\label{Fig2VQSPFramework}
\end{figure}

The PQC $U(\boldsymbol\theta)$ consists of the rotation layers and entangler layers \cite{Kandala2017Hardware,Havlivcek2019Supervised,Commeau2020Variational}, as shown in Fig. 2(a). Single qubit rotation layers include the rotation operator $R_{y}(\theta)=e^{-\iota\frac{\theta}{2}\sigma_{y}}$ depending on the tunable parameter $\theta$ and the Pauli operator $\sigma_{y}$. The two-qubit entangler layers are CNOT gates $CX=|0\rangle\langle0|\otimes I_2+|1\rangle\langle1|\otimes\sigma_{x}$ applying on the neighbor qubits, where $I_2$ is the identity with size $2\times2$. If we apply $L$ layers, the total number of parameters in this structure is $dL$. We remark that the circuit structure is flexible and the entangling gate can also be replaced with unitary $CZ=|0\rangle\langle0|\otimes I_2+|1\rangle\langle1|\otimes\sigma_{z}$ or $CR_y(\theta)=|0\rangle\langle0|\otimes I_2+|1\rangle\langle1|\otimes R_y(\theta)$.

For a given parameter $\boldsymbol\theta$, the probability $P_{j}^{\boldsymbol\theta}$ of obtaining outcome $j$ is generated via measuring the given trial state multiple times. In general, $P_{j}^{\boldsymbol\theta}=\langle\Phi(\boldsymbol\theta)|M_j|\Phi(\boldsymbol\theta)\rangle$ denotes the expectation value of the operator $M_j$ which can be represented as a linear combination of the Pauli tensor products such as
\begin{align}
M_j&=|j\rangle\langle j|=\bigotimes_{i=1}^{d}|j_i\rangle\langle j_i|\nonumber\\
&=\frac{1}{D}\bigotimes_{i=1}^{d}[I_2+(-1)^{j_i}\sigma_z]=\frac{1}{D}\sum_{i=0}^{D-1}M_j^i,
\end{align}
where $j=j_1j_2\cdots j_i\cdots j_d$ is a binary string and $j_i\in\{0,1\}$. The probability is therefore obtained in terms of the weighted sum of $D$ expectation values, $P_{j}^{\boldsymbol\theta}=\frac{1}{D}\sum_{i=0}^{D-1}\langle\Phi(\boldsymbol\theta)|M_j^i|\Phi(\boldsymbol\theta)\rangle$. The error of estimating $\langle M_j\rangle$ is $\epsilon^{2}=\sum_{i=0}^{D-1}\textrm{Var}[M_j^i]/(D^2S_j^i)$ for $S_j^i$ measurements of estimating $\langle\Phi(\boldsymbol\theta)|M_j^i|\Phi(\boldsymbol\theta)\rangle$ \cite{Cerezo2021Variational,Bharti2022Noisy}. Evaluating the overall probability distribution $P^{\boldsymbol\theta}=(P_0^{\boldsymbol\theta},
\cdots,P_{D-1}^{\boldsymbol\theta})$ requires to calculate the expectation values of $D$ operators \cite{Cerezo2022Variational}. Thus, this strategy would be inefficient due to the exponentially large $D$. One of alternative methods is classical sampling. Probabilities can be estimated from frequencies of a finite number of measurements. Crucially, we utilize quantum computer to sample the values of $j$. Let $S$ be the total number of samples and $j_1,j_2,\cdots,j_{S}$ be a sequence of outcomes. Denote $S_{j}$ the number of result $j$. One needs to calculate $P_j^{\boldsymbol\theta}\approx S_{j}/S$. From the Hoeffding's inequality \cite{Hoeffding1963Probability}, the sampling number of $S$ has a lower bound $S\geq\mathcal{O}(x_{\textrm{min}}^{-2})$, where $x_{\textrm{min}}=\min_ix_i$, which means that our algorithm is more efficient for larger $x_{\textrm{min}}$ and sparse data. Another alternative method is adaptive informationally complete generalized measurements \cite{Perez2021Learning}, which can be used to minimize statistical fluctuations.

The training of the PQC is achieved via optimizing the cost function
\begin{align}\label{CostFun1}
F_1(\boldsymbol\theta)=\left|\sqrt{D}\langle+^{\otimes d}|\Phi(\boldsymbol\theta)\rangle-\sum_{j=0}^{D-1}x_{j}\right|+F_1^{\textrm{KL}}(\boldsymbol\theta),
\end{align}
where the plus state $|+\rangle=(|0\rangle+|1\rangle)/\sqrt{2}$. The second term
\begin{align}
F_1^{\textrm{KL}}(\boldsymbol\theta)=-\sum_{j=0}^{D-1}x_{j}^{2}\log\frac{P_{j}^{\boldsymbol\theta}}{x_{j}^{2}}
\end{align}
is dubbed as the Kullback-Leibler divergence (KL) which quantifies the amount of information lost when changing from the probability distribution $x^{[2]}=(x_{0}^{2},\cdots,x_{D-1}^{2})$ to another distribution $P^{\boldsymbol\theta}$. The first term ensures the obtained states learned by the quantity $F_1^{\textrm{KL}}(\boldsymbol\theta)$ have positive local phases. For instance, we aim to prepare a single qubit state $|x\rangle=\frac{1}{\sqrt{5}}|0\rangle+\frac{2}{\sqrt{5}}|1\rangle$ corresponding to a classical vector $\vec{x}=(\frac{1}{\sqrt{5}},\frac{2}{\sqrt{5}})$. Variational optimizing the cost function $F_1^{\textrm{KL}}(\boldsymbol\theta)$ generates four states \cite{Zhu2019Training}
\begin{align}
&|x_{0}\rangle=\frac{1}{\sqrt{5}}|0\rangle+\frac{2}{\sqrt{5}}|1\rangle,~
|x_{1}\rangle=-\frac{1}{\sqrt{5}}|0\rangle+\frac{2}{\sqrt{5}}|1\rangle,\nonumber\\
&|x_{2}\rangle=\frac{1}{\sqrt{5}}|0\rangle-\frac{2}{\sqrt{5}}|1\rangle,~
|x_{3}\rangle=-\frac{1}{\sqrt{5}}|0\rangle-\frac{2}{\sqrt{5}}|1\rangle.\nonumber
\end{align}
Optimizing the first term can filter out the good state $|x_{0}\rangle$. It is clear that $F_1(\boldsymbol\theta)\geq0$ and the equality is true if and only if $U(\boldsymbol\theta_{\textrm{opt}})|0^{\otimes d}\rangle=|x\rangle$. Notice that the inner product $\langle+^{\otimes d}|\Phi(\boldsymbol\theta)\rangle$ is computed via the Hadamard test \cite{Aharonov2009A}.

Updating the circuit parameter by quantum-classical optimization loops, we produce an optimal parameter $\boldsymbol\theta_{\textrm{opt}}$ until the cost function converges towards zero. The learning scheme maps the classical vector into a set of parameters $\boldsymbol\theta_{\textrm{opt}}$, $\vec{x}\mapsto\{\boldsymbol\theta_{\textrm{opt}}\}$. Loading the parameter $\boldsymbol\theta_{\textrm{opt}}$ into NISQ devices equipped with a PQC $U(\boldsymbol\theta)$, we then prepare the state $|x\rangle\approx U(\boldsymbol\theta_{\textrm{opt}})|0^{\otimes d}\rangle$.

\emph{Case 2: an arbitrary real value vector.} In case 2, we consider a normalized vector $\vec{x}=(x_0,\cdots,x_j,\cdots,x_{D-1})$, $x_j\in\mathbb{R}$. Denote the position index set of nonnegative and negative elements by the subset $\mathscr{D}_{+}=\{k_1,k_2,\cdots,k_a\}\subset\mathscr{D}=\{0,1,\cdots,D-1\}$ and $\mathscr{D}_{-}=\{l_1,l_2,\cdots,l_b\}\subset\mathscr{D}$ in which $a+b=D$, $\mathscr{D}_{+}\cap\mathscr{D}_{-}=\emptyset$, and $\mathscr{D}_{+}\cup\mathscr{D}_{-}=\mathscr{D}$. We decompose the target state $|x\rangle=\sum_{j=0}^{D-1}x_j|j\rangle$ as a sum of two unnormalized states $|x_{+}\rangle$ and $|x_{-}\rangle$,
\begin{align}
|x\rangle=\sum_{j=0}^{D-1}x_j|j\rangle=|x_{+}\rangle+|x_{-}\rangle,
\end{align}
where the amplitudes of states $|x_{+}\rangle$ and $|x_{-}\rangle$ are nonnegative and negative, respectively. More precisely, we have the following forms
\begin{align}
&|x_{+}\rangle=\sum_{k=k_1}^{k_a}x_k|k\rangle+\sum_{l=l_1}^{l_b}0|l\rangle,~~x_k\geq0,\\
&|x_{-}\rangle=\sum_{k=k_1}^{k_a}0|k\rangle+\sum_{l=l_1}^{l_b}x_l|l\rangle,~~x_l<0.
\end{align}
For example, considering a state $|x\rangle=a|0\rangle+b|1\rangle$, $a>0$, $b<0$, $a^2+b^2=1$, the right decomposition is given as $|x\rangle=|x_{+}\rangle+|x_{-}\rangle$, where states $|x_+\rangle=a|0\rangle+0|1\rangle$ and $|x_-\rangle=0|0\rangle+b|1\rangle$.

By inserting a single ancillary qubits, we define a $(d+1)$-qubit state
\begin{align}
\label{X1}
|\Phi_{x}\rangle=|0\rangle|x_{+}\rangle-|1\rangle|x_{-}\rangle.
\end{align}
Next, we apply the Hadamard gate $\textrm{H}$ on the first qubit of state $|\Phi_{x}\rangle$ and yield
\begin{align}
|\Phi_{x}^{'}\rangle&=\left(\textrm{H}\otimes I_2^{\otimes d}\right)|\Phi_{x}\rangle\nonumber\\
&=|0\rangle\frac{|x_{+}\rangle-|x_{-}\rangle}{\sqrt{2}}+|1\rangle\frac{|x_{+}\rangle+|x_{-}\rangle}{\sqrt{2}}.
\end{align}
When we see the result $|1\rangle$ via measuring the first qubit, the state $|x\rangle$ is prepared with a success probability $1/2$.

Here, we prepare state $|\Phi_{x}\rangle$ (Eq. \ref{X1}) via Algorithm 1. The procedure starts with an initial state $|0^{\otimes(d+1)}\rangle$ and then prepares a trial state $|\Phi(\boldsymbol{\theta})\rangle=U(\boldsymbol{\theta})|0^{\otimes(d+1)}\rangle$. Sampling each qubits in the computational basis, we collect a probability distribution $P^{\boldsymbol\theta}=(P_0^{\boldsymbol\theta},P_1^{\boldsymbol\theta},\cdots,P_{2D-1}^{\boldsymbol\theta})$ in which only $D$ elements are required. The new cost function
\begin{align}
F_2(\boldsymbol\theta)&=\left|\sqrt{2D}\langle+^{\otimes(d+1)}|\Phi(\boldsymbol\theta)\rangle-\sum_{j=0}^{D-1}|x_{j}|\right|
+F_2^{\textrm{KL}}(\boldsymbol\theta),
\end{align}
where the second term
\begin{align}
F_2^{\textrm{KL}}(\boldsymbol\theta)=-\sum_{k=k_1}^{k_a}(x_{k})^2\log\frac{P_{k}^{\boldsymbol\theta}}{(x_{k})^2}
-\sum_{l=l_1}^{l_b}(x_{l})^2\log\frac{P_{l+D}^{\boldsymbol\theta}}{(x_{l})^2}.
\end{align}
Optimizing $F_2(\boldsymbol\theta)$ via classical optimization algorithms, we obtain an approximation state $|\Phi_{x}\rangle\approx|\Phi(\boldsymbol{\theta}_{\textrm{opt}})\rangle=U(\boldsymbol{\theta}_{\textrm{opt}})|0^{\otimes(d+1)}\rangle$ which can then be used to prepare $|x\rangle$ after applying $\textrm{H}\otimes I_2^{\otimes d}$ on it and seeing result $|1\rangle$.

\emph{Case 3: an arbitrary vector.} In case 3, we consider a normalized vector
\begin{align}
\vec{x}=(x_0,\cdots,x_j,\cdots,x_{D-1}),x_j=x_j^{\textrm{re}}+\iota x_j^{\textrm{im}}\in\mathbb{C}.\nonumber
\end{align}
Denote the real and imaginary parts by $x_j^{\textrm{re}}$ and $x_j^{\textrm{im}}$. Define index sets
\begin{align}
&\mathscr{D}_{+}^{\textrm{re}}=\{k_1,k_2,\cdots,k_a\}\subset\mathscr{D},
\mathscr{D}_{-}^{\textrm{re}}=\{l_1,l_2,\cdots,l_b\}\subset\mathscr{D},\nonumber\\
&\mathscr{D}_{+}^{\textrm{im}}=\{r_1,r_2,\cdots,r_c\}\subset\mathscr{D},
\mathscr{D}_{-}^{\textrm{im}}=\{s_1,s_2,\cdots,s_d\}\subset\mathscr{D},\nonumber
\end{align}
where each elements indicates the position index. For instance, $\mathscr{D}_{+}^{\textrm{re}}$ denotes the position index set of nonnegative real part of vector $\vec{x}$. It is clear to see that $a+b=c+d=D$ and
\begin{align}
&\mathscr{D}_{+}^{\textrm{re}}\cap\mathscr{D}_{-}^{\textrm{re}}=\emptyset,
\mathscr{D}_{+}^{\textrm{re}}\cup\mathscr{D}_{-}^{\textrm{re}}=\mathscr{D},\nonumber\\
&\mathscr{D}_{+}^{\textrm{im}}\cap\mathscr{D}_{-}^{\textrm{im}}=\emptyset,
\mathscr{D}_{+}^{\textrm{im}}\cup\mathscr{D}_{-}^{\textrm{im}}=\mathscr{D}.
\end{align}
The quantum state $|x\rangle$ has a decomposition,
\begin{align}
|x\rangle&=\sum_{j=0}^{D-1}x_j|j\rangle=\sum_{j=0}^{D-1}x_j^{\textrm{re}}|j\rangle+\iota\sum_{j=0}^{D-1}x_j^{\textrm{im}}|j\rangle\nonumber\\
&=|x_{+}^{\textrm{re}}\rangle+|x_{-}^{\textrm{re}}\rangle+\iota\left(|x_{+}^{\textrm{im}}\rangle+|x_{-}^{\textrm{im}}\rangle\right),
\end{align}
where unnormalized states
\begin{align}
&|x_{+}^{\textrm{re}}\rangle=\sum_{k=k_1}^{k_a}x_k^{\textrm{re}}|k\rangle,
~~|x_{-}^{\textrm{re}}\rangle=\sum_{l=l_1}^{l_b}x_l^{\textrm{re}}|l\rangle,\nonumber\\
&|x_{+}^{\textrm{im}}\rangle=\sum_{r=r_1}^{r_c}x_r^{\textrm{im}}|r\rangle,
~~|x_{-}^{\textrm{im}}\rangle=\sum_{s=s_1}^{s_d}x_s^{\textrm{im}}|s\rangle.
\end{align}

By adding two ancillary qubits, we prepare a $(d+2)$-qubit state
\begin{align}
\label{X2}
|\Phi_x\rangle=|00\rangle|x_{+}^{\textrm{re}}\rangle-|01\rangle|x_{-}^{\textrm{re}}\rangle
+|10\rangle|x_{+}^{\textrm{im}}\rangle-|11\rangle|x_{-}^{\textrm{im}}\rangle,
\end{align}
which amplitudes are nonnegative. Then, we perform the Hadamard gate $\textrm{H}$ and phase gate $S$ on ancillary qubits and obtain state
\begin{align}
|\Phi_x^{'}\rangle&=(S\otimes\textrm{H}\otimes I_2^{\otimes d})|\Phi_x\rangle\nonumber\\
&=|00\rangle\frac{|x_{+}^{\textrm{re}}\rangle-|x_{-}^{\textrm{re}}\rangle}{\sqrt{2}}
+|01\rangle\frac{|x_{+}^{\textrm{re}}\rangle+|x_{-}^{\textrm{re}}\rangle}{\sqrt{2}}\nonumber\\
&+\iota|10\rangle\frac{|x_{+}^{\textrm{im}}\rangle-\iota|x_{-}^{\textrm{im}}\rangle}{\sqrt{2}}
+\iota|11\rangle\frac{|x_{+}^{\textrm{im}}\rangle+\iota|x_{-}^{\textrm{im}}\rangle}{\sqrt{2}}.
\end{align}
Applying the Hadamard gate $\textrm{H}$ to the first qubit, we have
\begin{align}
|\Phi_x^{''}\rangle&=\left(\textrm{H}\otimes I_2^{\otimes (d+1)}\right)|\Phi_x^{'}\rangle\nonumber\\
&=|00\rangle\frac{|x_{+}^{\textrm{re}}\rangle-|x_{-}^{\textrm{re}}\rangle+\iota\left(|x_{+}^{\textrm{im}}\rangle-|x_{-}^{\textrm{im}}\rangle\right)}{2}\nonumber\\
&+|01\rangle\frac{|x_{+}^{\textrm{re}}\rangle+|x_{-}^{\textrm{re}}\rangle+\iota\left(|x_{+}^{\textrm{im}}\rangle+|x_{-}^{\textrm{im}}\rangle\right)}{2}\nonumber\\
&+|10\rangle\frac{|x_{+}^{\textrm{re}}\rangle-|x_{-}^{\textrm{re}}\rangle-\iota\left(|x_{+}^{\textrm{im}}\rangle-|x_{-}^{\textrm{im}}\rangle\right)}{2}\nonumber\\
&+|11\rangle\frac{|x_{+}^{\textrm{re}}\rangle+|x_{-}^{\textrm{re}}\rangle-\iota\left(|x_{+}^{\textrm{im}}\rangle+|x_{-}^{\textrm{im}}\rangle\right)}{2}.\nonumber
\end{align}
Finally, measuring the first two ancillary state, we produce state $|x\rangle$ by seeing the results $|01\rangle$ with success probability $1/4$.

Hence, we utilize Algorithm 1 to generate state $|\Phi_x\rangle$ (Eq. \ref{X2}). Set an initial state $|0^{\otimes(d+2)}\rangle$ and prepare a trial state $|\Phi(\boldsymbol{\theta})\rangle=U(\boldsymbol{\theta})|0^{\otimes(d+2)}\rangle$. Sampling each qubits in the computational basis, we collect a probability distribution $P^{\boldsymbol\theta}=(P_0^{\boldsymbol\theta},P_1^{\boldsymbol\theta},\cdots,P_{4D-1}^{\boldsymbol\theta})$ in which only $D$ elements are required. The cost function is defined as
\begin{align}
F_{3}(\boldsymbol{\theta})&
=\left|\sqrt{4D}\langle+^{\otimes(d+2)}|\Phi(\boldsymbol{\theta})\rangle-\sum_{j=0}^{D-1}(|x_j^{\textrm{re}}|+|x_j^{\textrm{im}}|)\right|\nonumber\\
&+F_3^{\textrm{KL}}(\boldsymbol{\theta}),
\end{align}
where the second term
\begin{align}
&F_3^{\textrm{KL}}(\boldsymbol{\theta})
=-\sum_{k=k_1}^{k_a}(x_k^{\textrm{re}})^2\log\frac{P_j^{\boldsymbol\theta}}{\left(x_k^{\textrm{re}}\right)^2}
-\sum_{l=l_1}^{l_b}\left(x_l^{\textrm{re}}\right)^2\log\frac{P_{l+D}^{\boldsymbol\theta}}{\left(x_l^{\textrm{re}}\right)^2}\nonumber\\
&-\sum_{r=r_1}^{r_c}\left(x_r^{\textrm{im}}\right)^2\log\frac{P_{r+2D}^{\boldsymbol\theta}}{\left(x_r^{\textrm{im}}\right)^2}
-\sum_{s=s_1}^{s_d}\left(x_s^{\textrm{im}}\right)^2\log\frac{P_{s+3D}^{\boldsymbol\theta}}{\left(x_s^{\textrm{im}}\right)^2}.
\end{align}
Performing the quantum-classical optimization loops, we learn an optimal parameter $\boldsymbol{\theta}_{\textrm{opt}}$ and produce an approximation state $|\Phi_x\rangle\approx|\Phi(\boldsymbol{\theta}_{\textrm{opt}})\rangle=U(\boldsymbol{\theta}_{\textrm{opt}})|0^{\otimes(d+2)}\rangle$ which can be further used to prepare the desired state $|x\rangle$.

Case 2 and 3 generalize the result of \cite{Liang2022Improved} to an arbitrary vector. The construction of Eqs. (\ref{X1}) and (\ref{X2}) is motivated by the work \cite{Nakaji2022Approximate} in which only real-value vector is considered.

\subsection{Quantum-assisted quantum simulation of open quantum systems}\label{subsecIII}
Based on the quantum-assisted quantum algorithm and VQSP, we investigate its application in quantum simulation of open quantum systems (OQS). Our simulation approach integrates three important components: a Choi-Jamiolkowski isomorphism technique \cite{Havel2003Robust,Kamakari2022Digital,Schlimgen2022Quantum,Ramusat2021Quantum,Yoshioka2020Variational} which maps the Lindblad master equation into a stochastic Schr$\ddot{\textrm{o}}$dinger equation with a non-Hermitian Hamiltonian, a variational quantum state preparation (VQSP) subroutine (introduced in subsection \ref{subsecII}) for the ancillary state preparation, and a method presented by works \cite{Barenco1995Elementary,Berry2015Simulating} to implement the block diagonal unitary.

Within the Markovian approximation, the dynamics of the density matrix $\rho(t)$ of an open quantum system is described by the Lindblad master equation \cite{Lindbald1976On,Weimer2021Simulation}
\begin{align}\label{LME}
\dot{\rho}(t)=-\iota[\mathcal{H},\rho(t)]+\mathcal{D}\rho(t),\rho(t)\in\mathbb{C}^{N\times N},N=2^n,
\end{align}
where $t$ and $\mathcal{H}$ are the time and system Hamiltonian. The dissipative superoperators
\begin{align}
\mathcal{D}\rho(t)=\sum_{r}\Big[\hat{L}_r\rho(t)\hat{L}_r^{\dag}-\frac{1}{2}\{\hat{L}_r^{\dag}\hat{L}_r,\rho(t)\}\Big].
\end{align}
Each Lindblad jump operator $\hat{L}_r$ describes the coupling to the environment. Here, we assume a local Lindblad equation such that both $\mathcal{H}$ and $\hat{L}_r$ are written as the sums of the tensor products of at most a few local degrees of freedom of the system \cite{Lloyd1996Universal}. Under the Choi-Jamiolkowski isomorphism \cite{Havel2003Robust,Kamakari2022Digital,Schlimgen2022Quantum,Ramusat2021Quantum,Yoshioka2020Variational} (see supplementary material for details \ref{MethodDetails}), the Eq. (\ref{LME}) can be rewritten in form of stochastic Schr$\ddot{\textrm{o}}$dinger equation, $|\dot{\rho}(t)\rangle=\hat{\mathcal{H}}|\rho(t)\rangle$ with a $N^{2}\times N^{2}$ non-Hermitian Hamiltonian
\begin{align}\label{nonhermitian}
\hat{\mathcal{H}}&=-\iota(I_{N}\otimes\mathcal{H}-\mathcal{H}^{T}\otimes I_{N})\nonumber\\
&+\sum_{r}\Big(\hat{L}_r^{*}\otimes\hat{L}_r
-\frac{1}{2}I_{N}\otimes\hat{L}_r^{\dag}\hat{L}_r-\frac{1}{2}\hat{L}_r^{T}\hat{L}_r^{*}\otimes I_{N}\Big)
\end{align}
in a doubled Hilbert space where $\hat{L}_r^{*}$ and $\hat{L}_r^{T}$ denote the complex conjugate and transpose of operator $\hat{L}_r$, $I_{N}$ is a $N\times N$ identity.

We assume that $\hat{\mathcal{H}}$ is $k$ local and has a unitary decomposition $\hat{\mathcal{H}}=\sum_{j=0}^{J-1}b_{j}\hat{\mathcal{H}}_{j}$ with positive real numbers $b_{j}>0$ (see supplementary material for details \ref{MethodDetails}). We remark that the Hamiltonian $\mathcal{H}$ and the dissipative operators $\hat{L}_r$ can be expressed as a linear combination of $q_{\mathcal{H}}$ and $q_{\hat{L}_r}$ easily implementable unitary operators. In particular, $q_{\mathcal{H}}$ and $q_{\hat{L}_r}$ scales polynomially with the number of qubits, $\mathcal{O}(\mbox{poly}(\log N))$. As a result the quantity $J=\mathcal{O}(\mbox{poly}(\log N))$. For instance, the XY spin chain with Hamiltonian, $\mathcal{H}_{xy}=\sum_{j=1}^{n-1}\left(\sigma_x^{j}\sigma_x^{j+1}
+\sigma_y^{j}\sigma_y^{j+1}\right)$ $(n\geq2)$, has $2(n-1)$ unitary operators \cite{Gibbs2021Long}. The Heisenberg $n$-qubit chain with open boundaries, $\mathcal{H}=-J\sum_{j=1}^{n-1}\left(\sigma_x^{j}\sigma_x^{j+1}+\sigma_y^{j}\sigma_y^{j+1}+\sigma_z^{j}\sigma_z^{j+1}\right)-h\sum_{j=1}^{n}\sigma_z^{j}$, has $3(n-1)+n=4n-3$ unitary operators \cite{Bespalova2021Hamiltonian}. Note that the quantity $J$ scales polynomially with the system size, $J=\mathcal{O}(\mbox{poly}(N))$, for certain Hamiltonians such as quantum systems with sparse interactions. For a more general case such as in chemistry problems, $J$ scales exponentially with the system size, $J=\mathcal{O}(N^4)$.

Given an initial pure state $|\rho(0)\rangle$, the time evolved state at time $T$ is given by $|\rho(T)\rangle=\frac{\mathcal{U}(T)|\rho(0)\rangle}{\|\mathcal{U}(T)|\rho(0)\rangle\|_2}$ with $\mathcal{U}(T)=e^{\hat{\mathcal{H}}T}$ \cite{Weimer2021Simulation}. Divide the time $T$ into $N_{T}$ segments of length $\Delta t=T/N_{T}$. The non-unitary operator $\mathcal{U}(\Delta t)=e^{\hat{\mathcal{H}}\Delta t}$ can be approximated as \cite{Berry2015Simulating}
\begin{align}
\mathcal{U}(\Delta t)\approx I_{N^2}+\Delta t\hat{\mathcal{H}}+\frac{(\Delta t\hat{\mathcal{H}})^2}{2!}
\end{align}
with error $\mathcal{O}(\Delta t^3)$, where the Taylor series is truncated at order $2$. The accuracy can be improved by using higher order Taylor expansion. In particular, the error is $\mathcal{O}(\Delta t^{J+1})$ when it is truncated on the order $J$. In this work, we choose a sufficient small $\Delta t$ and adapt a first order Taylor expansion, $\mathcal{U}(\Delta t)=\mathcal{Q}(\Delta t)+\mathcal{O}(\Delta t^{2})$. Here, the operator
\begin{align}\label{DecompositionA}
\mathcal{Q}(\Delta t)=I_{N^2}+\Delta t\Bigg(\sum_{j=0}^{J-1}b_{j}\hat{\mathcal{H}}_{j}\Bigg)=\sum_{j=0}^{J}a_{j}\mathcal{Q}_{j},
\end{align}
where $\mathcal{Q}_{j}=\hat{\mathcal{H}}_{j}$, $a_{j}=b_{j}\Delta t$ for $j=0,1,\cdots,J-1$ and $\mathcal{Q}_{J}=I_{N^{2}}$, $a_{J}=1$. We set $J+1=2^{m}$ for an integer $m$. If $J+1$ is not a power of $2$, we can divide the identity $I_{N^{2}}$ into several sub-terms until $J+1=2^{m}$ for updated $J$. Thus, for one time step evolution, the updated state $|\rho(\Delta t)\rangle=\mathcal{Q}(\Delta t)|\rho(0)\rangle/\|\mathcal{Q}(\Delta t)|\rho(0)\rangle\|_2$. Based on the Eq. (\ref{DecompositionA}), we realize such implementation on a quantum computer via the linear combination of unitaries (LCU) \cite{Long2006General,Long2008Duality}. Running the following Algorithm 2 for $\tau=0,1,\cdots,N_{T}-1$, we achieve the simulation of OQS up to time $T$.

\emph{Algorithm 2: a universal quantum algorithm for simulating OQS.}

(1) Prepare an ancillary state, $|a\rangle=\sum_{j=0}^{J}\sqrt{\frac{a_{j}}{A}}|j\rangle=U_{a}|0^{\otimes m}\rangle$, where $A=\sum_{j=0}^{J}a_{j}$.

(2) Construct a unitary $\Lambda_{\mathcal{Q}}=\sum_{j=0}^{J}|j\rangle\langle j|\otimes\mathcal{Q}_{j}$ and implement $(U_{a}^{\dag}\otimes I_{N^{2}})\Lambda_{\mathcal{Q}}$ on state $|a\rangle|\rho(\tau\Delta t)\rangle$. The system state now is
\begin{align}
|\Psi(\tau\Delta t)\rangle=A^{-1}|0^{\otimes m}\rangle\mathcal{Q}(\Delta t)|\rho(\tau\Delta t)\rangle+|\Psi_{\bot}(\tau\Delta t)\rangle\nonumber,
\end{align}
where the ancillary space of state $|\Psi_{\bot}(\tau\Delta t)\rangle$ is orthogonal to $|0^{\otimes m}\rangle$.

(3) Using the projector operator $\mathcal{P}=|0^{\otimes m}\rangle\langle0^{\otimes m}|\otimes I_{N^{2}}$ to measure the ancillary qubit, the system state is then collapsed into the state $\left|\rho\left((\tau+1)\Delta t\right)\right\rangle$ with a probability $P(\tau+1)=A^{-2}\|\mathcal{Q}(\Delta t)|\rho(\tau\Delta t)\rangle\|_{2}^{2}$. Performing classical sampling \cite{Montanaro2015Quantum}, $\mathcal{O}(1/P(\tau+1))$ repetitions are sufficient to prepare the part state
\begin{align}
|0^{\otimes m}\rangle\frac{\mathcal{Q}(\Delta t)|\rho(\tau\Delta t)\rangle}{\|\mathcal{Q}(\Delta t)|\rho(\tau\Delta t)\rangle\|_{2}}
=|0^{\otimes m}\rangle\left|\rho\left((\tau+1)\Delta t\right)\right\rangle.
\end{align}
Notice that the global success probability of preparing state $|\rho(\tau\Delta t)\rangle$ is $P_{\textrm{suc}}(\tau)=\prod_{i=1}^{\tau}P(i)$ which is exponentially small for larger $\tau$. This property indicates that Algorithm 2 has an exponential measurement cost for long time evolution and is only efficient for short time evolution.

\subsubsection{Two quantum-assisted algorithms for simulating open quantum systems}
Algorithm 2 provides a method of simulating OQS on a FTQ devices. The main obstacle is the implementation of depth unitary operators $U_{a}$ and $\Lambda_{\mathcal{Q}}$. Based on the quantum-assisted quantum framework introduced in subsection \ref{subsecI}, we here present two quantum-assisted algorithms to reduce the circuit depth of Algorithm 2 by utilizing the NISQ technology.

The first step is to approximately prepare the $m$-qubit quantum state,
\begin{align}
|a\rangle=\sum_{j=0}^{J}\sqrt{\frac{a_{j}}{A}}|j\rangle=U_{a}|0^{\otimes m}\rangle,
\end{align}
which corresponding to a normalized vector $\vec{a}=(\sqrt{a_{0}/A},\cdots,\sqrt{a_{J}/A})$. Instead of a direct construction of $U_{a}$, we utilize Algorithm 1 to learn an optimal parameter $\boldsymbol{\theta}_{\textrm{opt}}$ and feed it into a NISQ device equipped with a PQC $U(\boldsymbol\theta)$ \cite{Kandala2017Hardware,Havlivcek2019Supervised,Commeau2020Variational} to produce the state $U(\boldsymbol\theta_{\textrm{opt}})|0^{\otimes m}\rangle\approx|a\rangle$.

The second step is to compile the block diagonal unitary $\Lambda_{\mathcal{Q}}$ into a shallow quantum circuit. In the numerical example, we train a PQC $V(\boldsymbol{\beta})$ via optimizing the cost function
\begin{align}\label{FunctionG}
G(\boldsymbol{\beta})=1-\frac{1}{4^{2n+m}}\left|\textrm{Tr}(V^{\dag}(\boldsymbol{\beta})\Lambda_{\mathcal{Q}})\right|^2
\end{align}
on the parameter space to find the optimal parameter $\boldsymbol{\beta}_{\textrm{opt}}$ such that $\Lambda_{\mathcal{Q}}\approx V(\boldsymbol{\beta}_{\textrm{opt}})$. Function (\ref{FunctionG}) can be evaluated via the Hilbert-Schmidt test \cite{Khatri2019Quantum}. Moreover, as pointed in \cite{Khatri2019Quantum}, $G(\boldsymbol{\beta})$ would require exponential calls on $V(\boldsymbol{\beta})$ which means $G(\boldsymbol{\beta})$ is exponential fragile for high dimension system. Based on the above analysis, we propose two quantum-assisted quantum algorithms (Algorithm 3 and 4) for the simulation of OQS.

\emph{Algorithm 3: quantum-assisted quantum simulation of OQS.}

(1) The initial state $|0^{\otimes m}\rangle|\rho(\tau\Delta t)\rangle$.

(2) The evolved unitary
\begin{align}
W(\boldsymbol{\theta}_{\textrm{opt}},\boldsymbol{\beta}_{\textrm{opt}})
=\left(U^{\dag}(\boldsymbol{\theta}_{\textrm{opt}})\otimes I_{N^2}\right)
V(\boldsymbol{\beta}_{\textrm{opt}})\left(U(\boldsymbol{\theta}_{\textrm{opt}})\otimes I_{N^2}\right).
\end{align}

(3) Applying a measurement $\mathcal{P}$ on state
\begin{align}
|\Psi(\tau\Delta t)\rangle=W(\boldsymbol{\theta}_{\textrm{opt}},\boldsymbol{\beta}_{\textrm{opt}})|0^{\otimes m}\rangle|\rho(\tau\Delta t)\rangle,\nonumber
\end{align}
we obtain $|0^{\otimes m}\rangle|\rho((\tau+1)\Delta t)\rangle$ with a probability $P(\tau+1)$.

Motivated by the works \cite{Wei2020A}, Algorithm 4 starts with a different ancillary state
\begin{align}
|a^{'}\rangle=\sum_{j=0}^{J}\frac{a_{j}}{\sqrt{A^{'}}}|j\rangle=U^{'}(\boldsymbol{\theta}_{\textrm{opt}})|0^{\otimes m}\rangle,
A^{'}=\sum_{j=0}^{J}a_{j}^{2},
\end{align}
where $U^{'}(\boldsymbol{\theta}_{\textrm{opt}})$ learned by VQSP is a shallow quantum circuit. We then replace the unitary $W(\boldsymbol{\theta}_{\textrm{opt}},\boldsymbol{\beta}_{\textrm{opt}})$ with a new unitary
\begin{align}
W^{'}(\boldsymbol{\beta}_{\textrm{opt}})=(\textrm{H}^{\otimes m}\otimes I_{N^{2}})V(\boldsymbol{\beta}_{\textrm{opt}})\left(U^{'}(\boldsymbol{\theta}_{\textrm{opt}})\otimes I_{N^2}\right)
\end{align}
which utilizes $m$ Hadamard gate $\textrm{H}$.

\emph{Algorithm 4: an improved version of Algorithm 3.}

(1) The initial state $|0^{\otimes m}\rangle|\rho(\tau\Delta t)\rangle$.

(2) The evolved unitary $W^{'}(\boldsymbol{\beta}_{\textrm{opt}})$. The system state is
\begin{align}
|\Psi^{'}(\tau\Delta t)\rangle&=W^{'}(\boldsymbol{\beta}_{\textrm{opt}})|0^{\otimes m}\rangle|\rho(\tau\Delta t)\rangle\nonumber\\
&=\frac{1}{\sqrt{2^{m}A^{'}}}\sum_{j,j^{'}=0}^{J}a_{j}(-1)^{j\cdot j^{'}}|j^{'}\rangle\otimes\mathcal{Q}_{j}|\rho(\tau\Delta t)\rangle,\nonumber
\end{align}
where $j\cdot j^{'}$ denotes the bitwise inner product of $j,j^{'}$ modulo $2$.

(3) Apply a measurement $\mathcal{P}$ on state $|\Psi^{'}(\tau\Delta t)\rangle$. The system state is collapsed into $|0^{\otimes m}\rangle|\rho((\tau+1)\Delta t)\rangle$ with a probability $P^{'}(\tau+1)=\frac{\|\mathcal{Q}|\rho(\tau\Delta t)\rangle\|_{2}^{2}}{(J+1)A^{'}}$.

Compared with Algorithm 3, Algorithm 4 removes the unitary $U^{\dag}(\boldsymbol{\theta}_{\textrm{opt}})$. The Hadamard gates enable us to collect the states $\mathcal{Q}_{j}|\rho(\tau\Delta t)\rangle$ on the basis state $|j\rangle$ without the implementation of $U^{\dag}(\boldsymbol{\theta}_{\textrm{opt}})$. Let the error induced by unitary operators $U^{'}(\boldsymbol{\theta}_{\textrm{opt}})$($U(\boldsymbol{\theta}_{\textrm{opt}})$) and $V(\boldsymbol{\beta}_{\textrm{opt}})$ are $\epsilon_{\textrm{in}}$($\epsilon_{\textrm{in}}$) and $\epsilon_{\textrm{e}}$. The result error of Algorithm 4 roughly is $\epsilon_{\textrm{in}}+\epsilon_{\textrm{e}}$ lower than $2\epsilon_{\textrm{in}}+\epsilon_{\textrm{e}}$ obtained from Algorithm 3. Notice that the probability $P(\tau+1)\geq P^{'}(\tau+1)$ is true since the inequality $A^{2}\leq(J+1)A^{'}$ such that
\begin{align}
\Bigg[\sum_{j=0}^{J}a_{j}\Bigg]^2-(J+1)\sum_{j=0}^{J}a_{j}^2&=-\sum_{j=0}^{J}\sum_{k=j+1}^{J}(a_j-a_k)^2\leq0.
\end{align}
This means that Algorithm 4 has a lower success probability compared with Algorithm 3 and thus requires larger sampling complexity. Hence, Algorithm 4 further reduces the depth of Algorithm 3 but increase the sampling complexity.

\subsubsection{The measurement protocols}
Aside from simulating open quantum systems on a quantum computer, calculating the expectation value of an observable is also a particularly significant topic. Considering an observable $\mathcal{M}$, the expectation value $\langle\mathcal{M}\rangle=\textrm{Tr}[\mathcal{M}\rho(t)]$ can be estimated via an alternative, $\langle\mathcal{M}\rangle=\frac{\langle I_{N}|\hat{\mathcal{M}}|\rho(t)\rangle}{\langle I_{N}|\rho(t)\rangle}$, where the pure state $|I_{N}\rangle=\sum_{j=0}^{N-1}\frac{1}{\sqrt{N}}|jj\rangle$ is the vectorization of density operator $I_{N}$, $|j\rangle$ are the computational basis states and the operator $\hat{\mathcal{M}}=I_{N}\otimes\mathcal{M}$. The denominator ${\langle I_{N}|\rho(t)\rangle}$ ensures that the density matrix $\rho(t)$ is normalized. Suppose $\mathcal{M}$ can be efficiently encoded in terms of unitary operators $\mathcal{M}_j$ with coefficients $w_j$, $\mathcal{M}=\sum_jw_j\mathcal{M}_j$. In this subsection, we present two measurement protocols to calculate the expectation value.

\emph{Protocol A}. The first measurement protocol motivated by a recent work \cite{Kamakari2022Digital} relies on the Hadamard test \cite{Aharonov2009A}, which can be used to estimate the numerator and denominator of the expectation value. Given two unitary operators $\mathcal{U}_{I}$ (see Fig. 3.a) and $\mathcal{U}_{|\rho(t)\rangle}$ such that $\mathcal{U}_{I}|0^{\otimes2n}\rangle=|I_{N}\rangle$ and $\mathcal{U}_{|\rho(t)\rangle}|0^{\otimes2n}\rangle=|\rho(t)\rangle$. One performs a controlled unitary $|0\rangle\langle0|\otimes\mathcal{U}_{I}+|1\rangle\langle1|\otimes\mathcal{U}_{|\rho(t)\rangle}$ on an initial state $\frac{1}{\sqrt{2}}(|0\rangle+|1\rangle)|0^{\otimes2n}\rangle$ to obtain a state $\frac{1}{\sqrt{2}}(|0\rangle|I_{N}\rangle+|1\rangle|\rho(t)\rangle)$. After implementing a Hadamard gate $\textrm{H}$ on the ancillary qubit, we measure the qubit in the basis of Pauli operator $\sigma_{z}$ \cite{LiangWeiFei2022Quantum}. The real part of inner product $\langle I_{N}|\rho(t)\rangle$ is $2P_{0}-1$, where $P_{0}$ denotes the probability of the measurement outcome $|0\rangle$. Given accuracy $\epsilon$ with success probability at least $1-\delta$, the sampling complexity scales is $\mathcal{O}(\frac{1}{\epsilon^{2}}\log\frac{1}{\delta})$ \cite{LiangWeiFei2022Quantum}. The numerator can be expressed as
\begin{align}
\langle I_{N}|\hat{\mathcal{M}}|\rho(t)\rangle&=\langle I_{N}|I_{N}\otimes\mathcal{M}|\rho(t)\rangle\nonumber\\
&=\sum_jw_j\langle I_{N}|I_{N}\otimes\mathcal{M}_j|\rho(t)\rangle\nonumber\\
&=\sum_jw_j\langle I_{N}|\hat{\mathcal{M}}_{j}|\rho(t)\rangle,\hat{\mathcal{M}}_{j}=I_N\otimes\mathcal{M}_{j}.
\end{align}
Each term $\langle I_{N}|\hat{\mathcal{M}}_{j}|\rho(t)\rangle$ is obtained by replacing the controlled unitary $|0\rangle\langle0|\otimes\mathcal{U}_{I}+|1\rangle\langle1|\otimes\mathcal{U}_{|\rho(t)\rangle}$ with a unitary $|0\rangle\langle0|\otimes\mathcal{U}_{I}+|1\rangle\langle1|\otimes\hat{\mathcal{M}}_{j}\mathcal{U}_{|\rho(t)\rangle}$. However, in our simulation the unitary $\mathcal{U}_{|\rho(t)\rangle}$ is unknown. Basically, the evolved state is
\begin{align}
|0^{\otimes m}\rangle|\rho(t)\rangle=
\frac{\mathcal{P}W(\boldsymbol{\theta}_{\textrm{opt}},\boldsymbol{\beta}_{\textrm{opt}})|0^{\otimes m}\rangle|\rho(t-\Delta t)\rangle}
{\|\mathcal{P}W(\boldsymbol{\theta}_{\textrm{opt}},\boldsymbol{\beta}_{\textrm{opt}})|0^{\otimes m}\rangle|
\rho(t-\Delta t)\rangle\|_{2}},\nonumber
\end{align}
It is clear that the operator $\mathcal{P}W(\boldsymbol{\theta}_{\textrm{opt}},\boldsymbol{\beta}_{\textrm{opt}})$ is a non-unitary operator, which can not be implemented in the Hadamard test. The target unitary $\mathcal{U}_{|\rho(t)\rangle}$ can be found via quantum state learning \cite{Lee2018Learning,Chen2021Variational} and $|\rho(t)\rangle\approx\mathcal{U}_{|\rho(t)\rangle}|0^{\otimes2n}\rangle$. Note that \emph{Protocol A} requires to reproduce the state $|\rho(t)\rangle$ with a shallow quantum circuit $\mathcal{U}_{|\rho(t)\rangle}$. It is worth remarking that if we measure the $\tau$-th state $|\rho(\tau\Delta t)\rangle$, $(\tau+2)$ PQCs are learned via NISQ technology. In this case, our algorithms require more quantum resources than the usual variational quantum simulation. We next give an alternative \emph{Protocol B} to avoid the learning of other $\tau$ shallow quantum circuits.

\begin{figure*}[ht]
\centering
\includegraphics[scale=0.5]{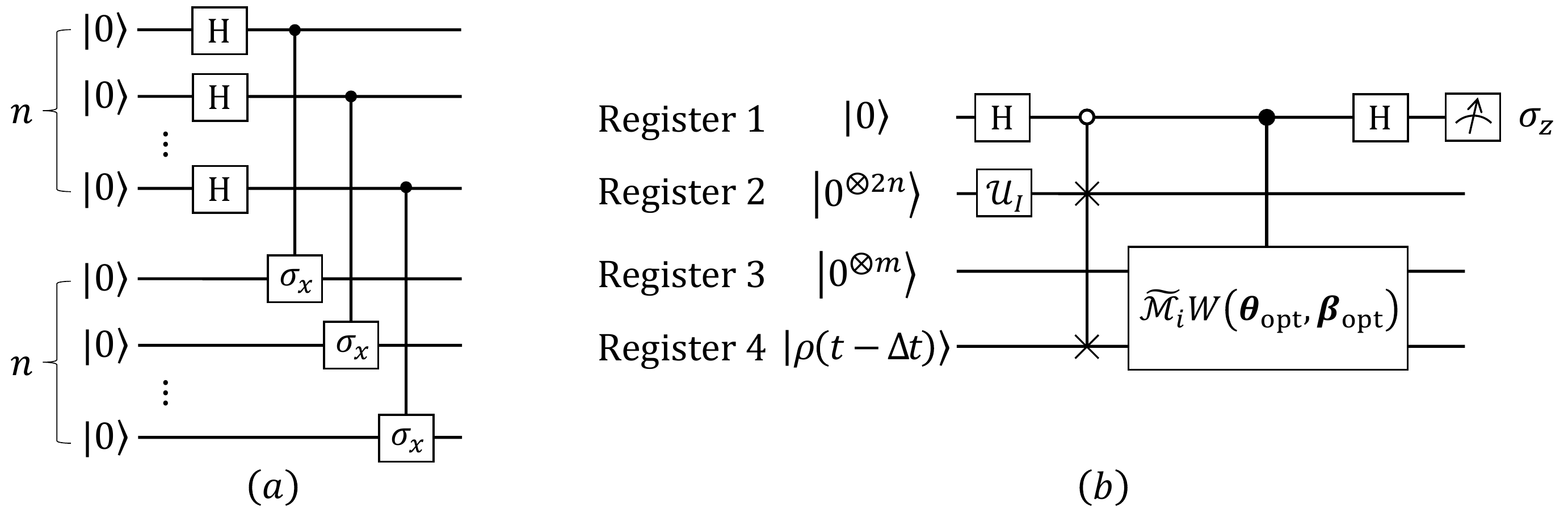}
\caption{Measurement circuits. (a) The circuit of $\mathcal{U}_{I}$. (b) Quantum circuit for estimating the real part of inner product $\langle0^{\otimes m}|\langle I_{N}|\widetilde{\mathcal{M}}_{i}|\Psi(t-\Delta t)\rangle$.}
\label{Fig3HadamardTest}
\end{figure*}

\emph{Protocol B}. The second measurement protocol removes the projector $\mathcal{P}$ after the implementation of the unitary $W(\boldsymbol{\theta}_{\textrm{opt}},\boldsymbol{\beta}_{\textrm{opt}})$ and does not produce the state $|\rho(t)\rangle$ itself. Instead, we consider a new observable consisting of some summation of local unitary operators $\widetilde{\mathcal{M}}_{i}$,
\begin{align}
\widetilde{\mathcal{M}}&=|0^{\otimes m}\rangle\langle0^{\otimes m}|\otimes\hat{\mathcal{M}}\nonumber\\
&=\frac{1}{J+1}\underbrace{(I_{2}+\sigma_{z})\otimes\cdots\otimes(I_{2}+\sigma_{z})}_{m}\otimes I_{N}\otimes\mathcal{M}\nonumber\\
&=\sum_{i}\tilde{w}_{i}\widetilde{\mathcal{M}}_{i}.
\end{align}
The expectation value $\langle\mathcal{M}\rangle$ is obtained in terms of the expectation value
\begin{align}\label{ExpectValue}
&\langle\mathcal{M}\rangle=\frac{\langle I_{N}|\hat{\mathcal{M}}|\rho(t)\rangle}{\langle I_{N}|\rho(t)\rangle}\nonumber\\
&=\frac{\langle0^{\otimes m}|\langle I_{N}|(I_{J+1}\otimes\hat{\mathcal{M}})\mathcal{P}|\Psi(t-\Delta t)\rangle}
{\langle0^{\otimes m}|\langle I_{N}|(I_{J+1}\otimes I_{N}\otimes I_{N})\mathcal{P}|\Psi(t-\Delta t)\rangle}\nonumber\\
&=\frac{\langle0^{\otimes m}|\langle I_{N}|(|0^{\otimes m}\rangle\langle0^{\otimes m}|\otimes\hat{\mathcal{M}})|\Psi(t-\Delta t)\rangle}
{\langle0^{\otimes m}|\langle I_{N}|(|0^{\otimes m}\rangle\langle0^{\otimes m}|\otimes I_{N}\otimes I_{N})|\Psi(t-\Delta t)\rangle}\nonumber\\
&=\frac{\sum_{i}\tilde{w}_{i}\langle0^{\otimes m}|\langle I_{N}|\widetilde{\mathcal{M}}_{i}|\Psi(t-\Delta t)\rangle}
{\langle0^{\otimes m}|\langle I_{N}|(|0^{\otimes m}\rangle\langle0^{\otimes m}|\otimes I_{N}\otimes I_{N})|\Psi(t-\Delta t)\rangle},
\end{align}
where $|\Psi(t-\Delta t)\rangle=W(\boldsymbol{\theta}_{\textrm{opt}},\boldsymbol{\beta}_{\textrm{opt}})|0^{\otimes m}\rangle|\rho(t-\Delta t)\rangle$.

Fig. 3(b) gives a quantum circuit for evaluating the numerator and denominator of Eq. (\ref{ExpectValue}). After applying unitary operators $\mbox{H}$ and $\mathcal{U}_{I}$ on registers 1,2 of state $|0\rangle|0^{\otimes2n}\rangle|0^{\otimes m}\rangle|\rho(t-\Delta t)\rangle$, we perform a controlled SWAP gate $|0\rangle\langle0|\otimes\mbox{SWAP}+|0\rangle\langle0|\otimes I_{N^2}\otimes I_{N^2}$ on registers 1,2,4 and obtain a state
\begin{align}
\frac{|0\rangle|\rho(t-\Delta t)\rangle|0^{\otimes m}\rangle|I_N\rangle+|1\rangle|I_N\rangle|0^{\otimes m}\rangle|\rho(t-\Delta t)\rangle}{\sqrt{2}}.
\end{align}
We next implement a controlled unitary $\widetilde{\mathcal{M}}_{i}W(\boldsymbol{\theta}_{\textrm{opt}},\boldsymbol{\beta}_{\textrm{opt}})$ on registers 1,3,4 and produce a state
\begin{align}
\frac{|0\rangle|\rho(t-\Delta t)\rangle|0^{\otimes m}\rangle|I_N\rangle+|1\rangle|I_N\rangle\widetilde{\mathcal{M}}_{i}|\Psi(t-\Delta t)\rangle}{\sqrt{2}}.
\end{align}
After performing a Hadamard gate $\mbox{H}$ on register 1, we measure register 1 and see result $|0\rangle$ with a probability
\begin{align}
P_0=\frac{1+\mbox{Re}\left(\langle\rho(t-\Delta t)|I_N\rangle\langle0^{\otimes m}|\langle I_N|\widetilde{\mathcal{M}}_{i}|\Psi(t-\Delta t)\rangle\right)}{2}.
\end{align}
If we perform a phase gate after the first Hadamard gate, the probability of seeing result $|0\rangle$ implying the imaginary part is
\begin{align}
P_0^{'}=\frac{1-\mbox{Im}\left(\langle\rho(t-\Delta t)|I_N\rangle\langle0^{\otimes m}|\langle I_N|\widetilde{\mathcal{M}}_{i}|\Psi(t-\Delta t)\rangle\right)}{2}.
\end{align}
Let $\langle\rho(t-\Delta t)|I_N\rangle=a+b\iota$ and $\langle0^{\otimes m}|\langle I_N|\widetilde{\mathcal{M}}_{i}|\Psi(t-\Delta t)\rangle=c+d\iota$ for real number $a,b,c,d$. We obtain a linear equation in terms of $P_0$ and $P_0^{'}$,
\begin{equation}\label{LinearEquation}
\left\{\begin{aligned}
& ac-bd &=2P_0-1, \\
& bc+ad &=1-2P_0^{'}.
\end{aligned}\right.
\end{equation}
Notice that the quantity $\langle\rho(t-\Delta t)|I_N\rangle$ has been estimated via the former process and therefore $a$ and $b$ are already known. Solving the Eq. (\ref{LinearEquation}), we can calculate $c$ and $d$ which are the real and imaginary parts of quantity $\langle0^{\otimes m}|\langle I_N|\widetilde{\mathcal{M}}_{i}|\Psi(t-\Delta t)\rangle$. Suppose the initial state $|\rho(0)\rangle=U_0|0^{\otimes 2n}\rangle$. The real and imaginary parts of $\langle\rho(0)|I_N\rangle=\langle0^{\otimes 2n}|U_0^{\dag}\mathcal{U}_I|0^{\otimes 2n}\rangle$ is estimated via the Hadamard test \cite{Aharonov2009A}.

\begin{table*}
\begin{tabular}{|c|c|c|}
\hline
 Algorithm &Number of qubits & Gate complexity \\ \hline
 Algorithm 2 &$2n+m$ & $\mathcal{O}[2^{m+1}+2^{m}km\log_{2}(\epsilon_{\mathcal{Q}}^{-1})]$ or $\mathcal{O}[2^{m+1}+2^{m}km^2]$ \\ \hline
 Algorithm 3 &$2n+m$ & $\mathcal{O}[2\mbox{poly}(m)+\mbox{poly}(2n+m)]$ \\ \hline
 Algorithm 4 &$2n+m$ & $\mathcal{O}[\mbox{poly}(m)+1+\mbox{poly}(2n+m)]$ \\ \hline
\end{tabular}
\caption{Resources scaling for different algorithms. $k$ is the locality of the non-Hermitian operator $\hat{\mathcal{H}}$ and $\epsilon_{\mathcal{Q}}$ are the error of implementing unitary $\Lambda_{\mathcal{Q}}$. The size of ancillary state is $2^{m}\times2^{m}$. The gate complexity denotes the number of quantum gates for small step $\Delta t$.}
\label{Comparison}
\end{table*}

\subsubsection{Circuit and measurement overhead of the simulation of OQS}
We here discuss the complexity of the quantum circuit, the number of required qubits and the number of measurements. For an $N\times N$ $(N=2^{n})$ Hamiltonian $\mathcal{H}$, $2n$-qubits are required to store the pure state $|\rho(0)\rangle$. The ancillary state $|a\rangle$ is stored on a $m$-qubit register. The total number of qubits needed is $T_{q}=2n+m$.

Without loss of generality we focus on one time step, $|\rho(0)\rangle\rightarrow|\rho(\Delta t)\rangle$. Algorithm 2 performs unitary $U_a$ and a block diagonal unitary $\Lambda_{\mathcal{Q}}$. Algorithms 3 and 4 implement two PQCs $U(\boldsymbol{\theta}_{\textrm{opt}})$, $V(\boldsymbol{\beta}_{\textrm{opt}})$ and Hadamard gate $\mbox{H}$. An exact quantum algorithm for state preparation would need at most $\mathcal{O}(2^{m})$ basic operations \cite{Plesch2011Quantum}. For a hardware-efficient Ansatz, $U(\boldsymbol{\theta}_{\textrm{opt}})$ contains $mL_{a}$ single qubit unitaries $R_{y}(\theta)=e^{-\iota\theta\sigma_{y}/2}$ with parameter angle $\theta$, where $L_{a}$ is the depth of unitary $U(\boldsymbol{\theta}_{\textrm{opt}})$. Thus the cost of implementing $U(\boldsymbol{\theta}_{\textrm{opt}})$ is $\mathcal{O}(\mbox{poly}(m))$. Consequently, the VQSP subroutine achieves an important reduction on the gate complexity. Next, we quantify the overhead of implementing the block diagonal unitary $\Lambda_{\mathcal{Q}}$. In general, decomposing an arbitrary quantum circuit into a sequence of basic operations is a significant challenge \cite{Zhou2011Adding}. However, based on the locality of the non-Hermitian Hamiltonian $\hat{\mathcal{H}}$ ($k$-local), the cost of approximate and exact simulating the unitary gate $\Lambda_{\mathcal{Q}}$ within $\epsilon_{\mathcal{Q}}$ is $\mathcal{O}(2^{m}km\log_{2}(\epsilon_{\mathcal{Q}}^{-1}))$ and $\mathcal{O}(2^{m}km^2)$ (see supplementary material for details \ref{MethodDetails}). Crucially when $\Lambda_{\mathcal{Q}}$ is learned by a shallow quantum circuit $V(\boldsymbol{\beta}_{\textrm{opt}})$ with depth $L_{e}$, this part can be approximately simulated with $\mathcal{O}(\mbox{poly}(2n+m))$ quantum gates \cite{Khatri2019Quantum,Yu2022Optimal}. The depth of Algorithm 3 and 4 are $2L_{a}+L_{e}$ and $L_{a}+L_{e}$ independent on the system size. Table 1 compares circuit gate complexity of Algorithms 2, 3, and 4. It can be observed that Algorithm 3 and 4 reduce the gate complexity for small step $\Delta t$.

Another important aspect of assessing an algorithm is the measurement cost. In our approaches, the measurement overhead scales with the inversion of the success probability, $\mathcal{O}(1/P_{\textrm{suc}}(t))$. With the increase of the iterative steps, the success probability $P_{\textrm{suc}}(t)$ would decrease exponentially for larger time $t$ and hence induce exponential large measurement overhead. Thus, our approach is efficient for a short time evolution. For enough long time evolution, our approaches require exponential measurement cost. In next subsection, we numerically demonstrate this claim for different examples.

\subsubsection{Comparison with variational quantum simulation}
Variational quantum simulation (VQS) has been systematically studied in Ref. \cite{Yuan2019Theory}, which simulates a time evolution by learning a PQC at each small step $\Delta t$ \cite{Yao2021Adaptive}. The optimal PQC is obtained by minimizing the squared McLachlan distance between the variational and the exact states. These approaches require matrix inversion and the corresponding measurement cost scales $\mathcal{O}(\kappa^2\varepsilon^{-2})$ for accuracy $\varepsilon$, as well as the condition number $\kappa$ \cite{Benedetti2021Hardware}. The factor $\kappa^2$ poses computational challenges for ill-conditioned matrices. After $\tau$ small steps, the system state $|\rho(\tau\Delta t)\rangle=U_{\tau}U_{\tau-1}\cdots U_{1}|\rho(0)\rangle$ is generated via $\tau$ PQCs. The overall process is coherent and does not need to make measurement during the process. However, our algorithm follows the LCU framework and only two PQCs ($U(\boldsymbol{\theta}_{\textrm{opt}})$ and $V(\boldsymbol{\beta}_{\textrm{opt}})$) are learned for preparing the ancillary state and the evolution unitary, respectively. Each step is simulated by performing unitary $W(\boldsymbol{\theta}_{\textrm{opt}},\boldsymbol{\beta}_{\textrm{opt}})$, followed by a projective measurement $\mathcal{P}$. Repeating the process $\tau$ times, we obtain the state $|\rho(\tau\Delta t)\rangle$. We here assumed that $\tau>2$ and not to large such that the measurement probability of obtaining the state $|\rho(\tau\Delta t)\rangle$ is not exponentially small.

Compared with the VQS, our algorithm has two fold advantages. The first one is that our algorithm can evolve arbitrary initial state $|\rho(0)\rangle$ without learning new unitary except $U(\boldsymbol{\theta}_{\textrm{opt}})$ and $V(\boldsymbol{\beta}_{\textrm{opt}})$. However, the VQS requires to learn different unitary for different initial states. The second advantage is that our algorithm reduces the required number of PQCs even for a fixed initial state $|\rho(0)\rangle$ if $\tau>2$. Thus, we propose alternatives that do not require matrix inversion in simulating open quantum systems.

The error of our algorithm comes from two aspects. The first one is the truncation error of the evolution operator. This error can be reduced by taking small time step $\Delta t$ and performing high-order Taylor truncation. However, high-order Taylor truncation would increase the circuit complexity. The second one is the learning error for unitary operators $U(\boldsymbol{\theta}_{\textrm{opt}})$ and $V(\boldsymbol{\beta}_{\textrm{opt}})$. Smaller values of cost functions ($F_1(\boldsymbol{\theta})$ and $G(\boldsymbol{\beta})$) imply small errors. Similar to the general VQAs, high expressively ansatz and clear optimization method may yield small cost functions \cite{Cerezo2021Variational,Bharti2022Noisy}.

\subsection{Numerical results}
To exhibit our algorithms, we consider the dynamics given by a two-level system with an amplitude damping channel and a open version of the dissipative transverse field Ising model (DTFIM) on two sites \cite{Schlimgen2022Quantum}.
\begin{figure*}[ht]
\centering
\includegraphics[scale=0.8]{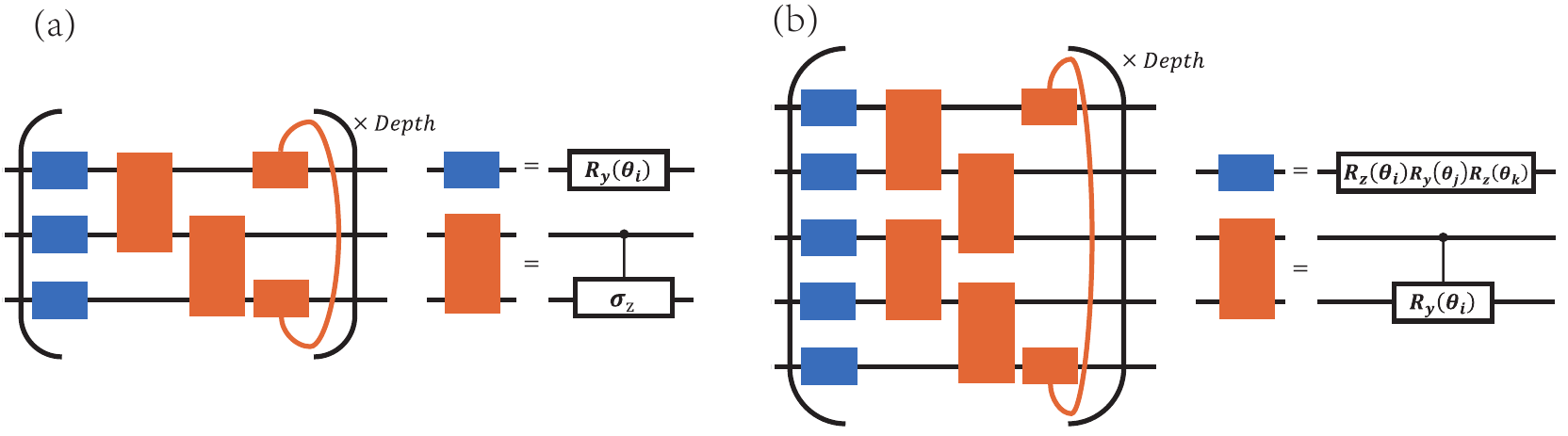}
\caption{PQCs $U(\boldsymbol{\theta})$ (a) and $V(\boldsymbol{\beta})$ (b) for preparing $|a\rangle$ and compiling unitary $\Lambda_{\mathcal{Q}}$, respectively.}
\label{Fig4PQCaq}
\end{figure*}

\begin{figure*}[ht]
\centering
\includegraphics[scale=0.55]{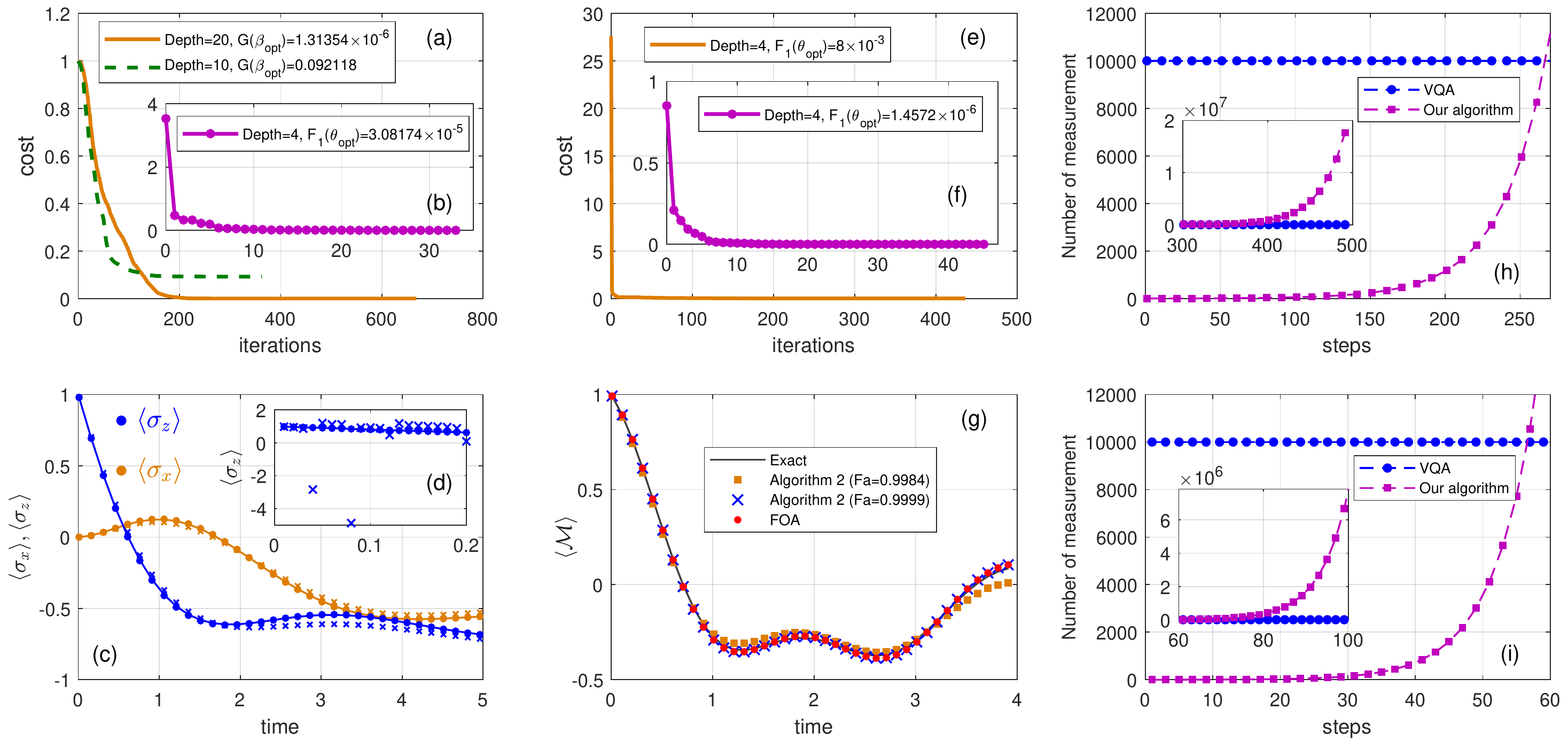}
\caption{Numerical results. (a,b,c,d) show the numerical results of Example 1. (a). The training process of compiling unitary $\Lambda_{\mathcal{Q}}$ with circuit depths $10$ and $20$. (b). The training process of preparing $|a\rangle$ with circuit depth $4$. (c and d). The dynamics of a two-level system with an amplitude damping channel with $\delta=\Omega=\gamma=1$ and small time $\Delta t=0.01$. The colors, orange and blue, corresponding to the expectation values of operators $\sigma_x$ and $\sigma_z$, respectively. The filled circles denote the simulation result of the first order approximation operator of evolution $e^{\hat{\mathcal{H}}\Delta t}$. The $\textrm{x}$ denotes the results of Algorithm 3 with $G(\boldsymbol{\beta}_{\textrm{opt}})=1.31354\times10^{-6}$ (c) and $G(\boldsymbol{\beta}_{\textrm{opt}})=0.092118$. (d). The solid lines presents the exact simulation. (e,f,g). The numerical results of Example 2. FOA represents the results of implementing the first order approximation of the non-unitary operator. The solid circuit and the red circle denote the exact and FOA simulation. The blue x and orange square indicate the result of Algorithm 2 with ancillary state fidelity $0.9999$ and $0.9984$. (h) and (i). The comparison of measurement complexity between standard VQAs and our quantum algorithm for example 1 (h) and 2 (i).}
\label{Fig5Results}
\end{figure*}

\emph{Example 1: A two-level system with an amplitude damping channel}. The Hamiltonian of a two-level system $\mathcal{H}=-\frac{\delta}{2}\sigma_z-\frac{\Omega}{2}\sigma_x$ and the Lindblad operator is an amplitude damping channel $\hat{L}=\frac{\sqrt{\gamma}}{2}(\sigma_x-\iota\sigma_y)$. We also have the equations $\hat{L}^{*}=\hat{L}=\frac{\sqrt{\gamma}}{2}(\sigma_x-\iota\sigma_y)$ and $\hat{L}^{T}=\hat{L}^{\dag}=\frac{\sqrt{\gamma}}{2}(\sigma_x+\iota\sigma_y)$. Based on the Eq. (\ref{nonhermitian}), we have the non-Hermitian Hamiltonian
\begin{align}
\hat{\mathcal{H}}&=-\iota(I_2\otimes\mathcal{H}-\mathcal{H}^{T}\otimes I_2)\nonumber\\
&+(\hat{L}^{*}\otimes\hat{L}-\frac{1}{2}I_2\otimes\hat{L}^{\dag}\hat{L}-\frac{1}{2}\hat{L}^{T}\hat{L}^{*}\otimes I_2).
\end{align}
Hence, the first order truncation of the operator $e^{\hat{\mathcal{H}}\Delta t}$ is $\mathcal{Q}(\Delta t)=\sum_{j=0}^{7}a_j\mathcal{Q}_j$. The coefficients $a_j$ and unitary operators $\mathcal{Q}_j$ are shown in Table 2.

Based on VQSP, we first train a $3$-qubit PQC $U(\boldsymbol{\theta})$ shown in Fig. 4(a) to prepare the ancilla state $|a\rangle$ which corresponding to an unnormalized vector of classical vector $\vec{a}$ of the form
\begin{align}
\left(\frac{\Delta t\delta\sqrt{2}}{2},\frac{\Delta t\gamma}{4},\sqrt{\alpha^2+\beta^2},
\frac{\Delta t\gamma}{4},\frac{\Delta t\gamma}{4},\frac{\Delta t\gamma}{4},\frac{\Delta t\gamma}{4},\frac{\Delta t\gamma}{4}\right).
\end{align}
The classical optimization of parameters $\boldsymbol{\theta}$ is achieved via using the Broyden-Fletcher-Goldfarb-Shanno (BFGS) quasi-Newton algorithm \cite{Broyden1970The,Fletcher1970A,Goldfarb1970A,Shanno1970Conditioning}. Fig. 5(b) shows the training process with the circuit depth $4$ and the final cost function value $F(\boldsymbol{\theta}_{\textrm{opt}})=3.08174\times10^{-5}$. Next, we compile the block diagonal unitary $\Lambda_{\mathcal{Q}}=\sum_{j=0}^{7}|j\rangle\langle j|\otimes\mathcal{Q}_j\in\mathbb{C}^{2^5\times2^5}$ into a $5$-qubit shallow quantum circuit $V(\boldsymbol{\beta})$ shown in Fig. 4(b). The cost function is defined by \cite{Khatri2019Quantum}
\begin{align}
G(\boldsymbol{\beta})=1-\frac{\left|\textrm{Tr}\left(V^{\dag}(\boldsymbol{\beta})\Lambda_{\mathcal{Q}}\right)\right|^2}{4^{5}},
\end{align}
which can be evaluated via the Hilbert-Schmidt test \cite{Khatri2019Quantum}. We find from Fig. 5(a) that the lower depth (depth$=10$) has a worse performance $(G(\boldsymbol{\beta}_{\textrm{opt}})=0.092118)$ for compiling unitary $\Lambda_{\mathcal{Q}}$ compared with larger depth $=20$ $(G(\boldsymbol{\beta}_{\textrm{opt}})=1.31354\times10^{-6})$. Fig. 5(c and d) show the dynamics with initial state $|\rho(0)\rangle=|00\rangle$ for different compiling precisions. It is clear to see that higher compiling precision has better performance.
From Fig. 5(c), we see that our simulation results are in agreement with the exact results for the former $200$ steps (time $t\leq2$). For more larger evolution time, our approach is unable to match the first order approximation (FOA) simulation exactly. The reasons are that the learned unitary operators are not exact. In order to increase the precision, one of strategies is to construct high expressively ansatz. This can be achieved by designing a larger parameter space so as to cover the exact unitary operators as well as possible.
\begin{table*}
\begin{center}
\label{Example1LCU}
\begin{tabular}{|c|c|c|c|c|c|c|}
\hline
$j$ & 0 & 1 & 2 & 3 & 4 & 5 \\
\hline
$a_j$ & $\frac{\sqrt{2}\Delta t\delta}{2}$ & $\frac{\gamma\Delta t}{4}$ & $\sqrt{\alpha^2+\beta^2}$ & $\frac{\gamma\Delta t}{4}$ & $\frac{\gamma\Delta t}{4}$ & $\frac{\gamma\Delta t}{4}$ \\
\hline
$\mathcal{Q}_j$ & $I_2^{(1)}\otimes\iota\textrm{H}^{(2)}$ & $I_2^{(1)}\otimes\sigma_z^{(2)}$ & $\frac{-\iota\alpha\textrm{H}^{(1)}+\beta I_2^{(1)})}{\sqrt{\alpha^2+\beta^2}}$
& $-\sigma_{z}^{(1)}\otimes I_2^{(2)}$ & $\sigma_{x}^{(1)}\otimes \sigma_{x}^{(2)}$ & $\sigma_{x}^{(1)}\otimes -\iota\sigma_{y}^{(2)}$ \\
\hline
$j$ & 6 & 7 & & & &  \\
\hline
$a_j$ & $\frac{\gamma\Delta t}{4}$ & $\frac{\gamma\Delta t}{4}$ & & & &\\
\hline
$\mathcal{Q}_j$ & $-\iota\sigma_{y}^{(1)}\otimes\sigma_{x}^{(2)}$ & $-\sigma_{y}^{(1)}\otimes\sigma_{y}^{(2)}$ & & & &\\
\hline
\end{tabular}
\end{center}
\caption{The coefficients $a_j$ and unitary operators $\mathcal{Q}_j$. $\textrm{H}$ denotes the Hadamard gate and $\alpha=\frac{\sqrt{2}\delta\Delta t}{2}$, $\beta=\left(1-\frac{\gamma\Delta t}{2}\right)$.}
\end{table*}

\emph{Example 2: The dissipative transverse field Ising model.} The Hamiltonian of the DTFIM with $2$ sites is $\mathcal{H}=-V\sigma_{z}^{(1)}\sigma_{z}^{(2)}-\Omega\sum_{k=1}^{2}\sigma_{x}^{(k)}$ and the Lindblad operators $\sqrt{\gamma}\sigma_{-}^{(k)}=\sqrt{\gamma}(\sigma_{x}^{(k)}-\iota\sigma_{y}^{(k)})$ with decay rate $\gamma$, interaction strength $V$ between adjacent spins and transverse magnetic field $\Omega$.

The first order approximation operator $I_{16}+\hat{\mathcal{H}}\Delta t$ can be decomposed into $32$ unitary operators with real positive coefficients (see supplementary material for details \ref{MethodDetails}). Thus, $5$ qubits are required to prepare the ancilla state $|a\rangle$. Figs. 5(.e and f) show the training process of learning $U(\boldsymbol{\theta}_{\textrm{opt}})$ with different PQCs. The $5$-qubit PQC $U(\boldsymbol{\theta})$ has a similar structure like Fig. 4(b). But the single qubit unitary of Fig. 5(e) is $R_{y}(\theta_i)$ rather than $R_{z}(\theta_i)R_{y}(\theta_j)R_{z}(\theta_k)$. In Fig. 5(f), the single qubit unitary also is $R_{y}(\theta_i)$ and the two qubit unitary is set as a controlled NOT gate rather a controlled unitary $R_{y}(\theta_i)$. In Figs. 5(e and f), the fidelity of preparing $|a\rangle$ are $0.9984$ and $0.9999$, respectively. Fig. 5(g) shows the average magnetization of the DTFIM, $\frac{1}{2}\sum_{k}\langle\sigma_{z}^{(k)}\rangle$, with the initial state $|\rho(0)\rangle=|0000\rangle$ and $V=\Omega=1$, $\gamma=0.1$. Fig. 5(g) indicates for the higher fidelity $0.9999$ the simulation approaches exact result much closer than lower fidelity $0.9984$. Notice that we follow the Algorithm 3 but exact implement the block-diagonal unitary $\Lambda_{\mathcal{Q}}$ in Example 2, since the rightness of compiling approaches have been verified in works \cite{Khatri2019Quantum,Yu2022Optimal}.

Finally, we compare the number of measurement samples between our algorithms and general VQAs. To achieve a certain precision $\varepsilon$, the standard sample complexity of VQAs scales $\mathcal{O}(\varepsilon^{-2})$ \cite{Cerezo2021Variational,Bharti2022Noisy}. However, the sample complexity of our algorithm as demonstrated in Algorithm 2 is the inversion of the success probability, $\mathcal{O}(P_{\textrm{suc}}^{-1})$. Here, we select $\varepsilon=10^{-2}$ and the corresponding number of measurement is $10^{4}$. For small time steps ($255$ or $55$ steps) as shown in Figs. 5(h and i), the sample complexity of our algorithm is lower than VQAs. With the increasing of the time steps, the sample complexity of our algorithm is exponential large as shown in subfigure of Figs. 5(h and i). Hence, our algorithm is more efficient for small time steps.

\section{Discussion}
We have introduced a quantum-assisted quantum algorithm framework to execute a UQA on NISQ devices. This framework not only follows the structure of the UQA but also reduces the circuit depth with the help of NISQ technology. To benefit from the advantages of NISQ technology maximally, we need to consider two essential issues, the NISQ preparation for general quantum states and the quantum compilation of unitary processes appeared in UQA. Our result bridges the technology gap between universal quantum computers and the NISQ devices by opening a new avenue to assess and implement a UQA on NISQ devices. Moreover, the framework emphasizes the important role of NISQ technologies beyond NISQ era. As shown in our work, NISQ technologies may be a useful tool for simulating open quantum systems.

Furthermore, we have developed a VQSP to prepare an amplitude encoding state of arbitrary classical data. The preparation algorithm generalized the result of the recent work \cite{Liang2022Improved} in which only positive classical data is considered. We have shown that any classical vector with size $2^d$ can be loaded into a $d$-qubit quantum state with a shallow depth circuit using only single- and two-qubit gates. The cost function to train the PQC is defined as a Kullback-Leibler divergence with a penalty term. It is the penalty term that filters the states with local phases. The maximal number of ancillary qubits is only two qubits in which the classical vector is a complex vector. In particular, no any other ancillary qubit is required
for an nonnegative vector and only single qubit for a real vector with local sign. This subroutine may have potential applications in quantum computation and quantum machine learning. However, the current method is more efficient for low-dimension and sparse data. For exponential dimension vector, our method would require exponential measurement cost like other works \cite{Zhu2019Training,Liang2022Improved}.

Based on the proposed framework and the VQSP, we have presented two quantum-assisted quantum algorithms for simulating an open quantum system with logarithmical quantum gate resources on limited time steps. The performance of our approach depends on the precision of the learned two unitary operators $U(\boldsymbol{\theta}_{\textrm{opt}})$ and $V(\boldsymbol{\beta}_{\textrm{opt}})$. For bigger sizes and more strongly correlated problems, learning two PQCs requires more quantum resources including the number of qubits and layers. This would increase the difficulty of learning optimal PQCs. Hence, our approach may get worse in practice for bigger size and more strongly correlated problems. Notice that our quantum-assisted quantum simulation framework can be naturally used to simulate general processes such as closed quantum systems \cite{Berry2015Simulating} and a quantum channel \cite{Nielsen2000Quantum} (see supplementary material for details \ref{MethodDetails}).

We here remark that although \emph{quantum-assisted quantum algorithm} may reduce the circuit depth of a UQA by learning two PQCs, whether the compiled quantum-assisted quantum algorithm are efficient and available is also related with the measurement cost. For instance, for long time evolution, the proposed quantum-assisted quantum simulation of open quantum systems would be inefficient because exponential measurement cost.

Recently, we note that the authors in \cite{Yalovetzky2022NISQ} studied the execution of the HHL algorithm on NISQ hardware. The two expensive components of the HHL on NISQ devices are the eigenvalue inversion and the preparation of right-side vector. The quantum phase estimation is replaced with the quantum conditional logic for estimating the eigenvalues. Nevertheless, the right-side vector $|b\rangle$ is assumed in advance to be similar to the HHL. Hence, a quantum-assisted HHL is possible by removing the assumption with the proposed VQSP in the algorithm \cite{Yalovetzky2022NISQ}.
\section*{Limitations of the study}
The first limitation is the uncontrollable of NISQ technology such as the expressibility of the chosen ansatz and the barren plateau \cite{Bharti2022Noisy,Cerezo2021Variational}. Another limitation is the measurement cost. Incompatible and partial measurement have recently been discussed in Refs. \cite{Long2021Collapse,Zhang2022A}. It would be interesting to mitigate the exponential compression of the success probability by combining the variational quantum simulation approaches \cite{Endo2020Variational,Benedetti2021Hardware}.

\section*{Acknowledgements}
This work is supported by the National Natural Science Foundation of China (NSFC) under Grant Nos. 12075159 and 12171044; Beijing Natural Science Foundation (Grant No. Z190005); the Academician Innovation Platform of Hainan Province.

Author contributions: J.M. and Q.Q. conceived the study. J.M., Q.Q., Z.X., and S.M. wrote the manuscript. All authors reviewed and critically revised the manuscript.

Declaration of interests: The authors declare no competing interests.

Materials Availability: This study did not generate new materials.

Data and code availability:

$\bullet$ Data reported in this paper will be shared by the lead contact upon request.

$\bullet$ This paper does not report original code.

$\bullet$ Any additional information required to reanalyze the data reported in this paper is available from the lead contact upon request.

\clearpage
\onecolumngrid
\section*{Supplementary Material}\label{MethodDetails}
\textbf{1. Choi-Jamiolkowski isomorphism.}

We first review the Choi-Jamiolkowski isomorphism \cite{Havel2003Robust} and utilize it to map the master equation to a stochastic Schr$\ddot{\textrm{o}}$dinger equation.

Given an arbitrary density matrix $\rho=\sum_{jk}\rho_{jk}|j\rangle\langle k|\in\mathbb{C}^{N\times N}$, its vectorized density operator (unnormalized) under Choi-Jamiolkowski isomorphism is given by $|\rho\rangle=\sum_{jk}\rho_{jk}|k\rangle\otimes|j\rangle$ in a doubled space with size $N^{2}\times N^{2}$. Here, we outline three useful properties.

(i) The vectorized form of an identity operator $I_{N}$ is $|I_{N}\rangle=\sum_{j}|j\rangle\otimes|j\rangle$.

(ii) The trace of a density matrix $\rho$ can be written in inner product form, $\textrm{Tr}(\rho)=\langle I_{N}|\rho\rangle$. In our simulation approach, $\textrm{Tr}(\rho)=1$ cannot be fulfilled. This means that the operator $\rho$ obtained from $|\rho\rangle$ is not a density matrix. Thus we require to normalize in estimating the expectation value of an observable.

(iii) For operators $A,B$ and a density operator $\rho$, we have $|A\rho B\rangle=(B^{T}\otimes A)|\rho\rangle$.

In this way, we have the following relations,
\begin{align}
|\mathcal{H}\rho(t)-\rho(t)\mathcal{H}\rangle&=(I_{N}\otimes\mathcal{H}-\mathcal{H}^{T}\otimes I_{N})|\rho(t)\rangle,
\nonumber\\
|\hat{L}_k\rho(t)\hat{L}_k^{\dag}\rangle&=\hat{L}_k^{*}
\otimes\hat{L}_k|\rho(t)\rangle,\nonumber\\
|\hat{L}_k^{\dag}\hat{L}_k\rho(t)+\rho(t)\hat{L}_k^{\dag}\hat{L}_k\rangle
&=(I_{N}\otimes\hat{L}_k^{\dag}\hat{L}_k+\hat{L}_k^{T}\hat{L}_k^{*}\otimes I_{N})|\rho(t)\rangle.\nonumber
\end{align}
The master equation can then be written as $|\dot{\rho}(t)\rangle=\hat{\mathcal{H}}|\rho(t)\rangle$, where
\begin{align}
\hat{\mathcal{H}}&=-\iota(I_{N}\otimes\mathcal{H}-\mathcal{H}^{T}\otimes I_{N})
+\sum_{k}\Big(\hat{L}_k^{*}\otimes\hat{L}_k-\frac{1}{2}I_{N}
\otimes\hat{L}_k^{\dag}\hat{L}_k-\frac{1}{2}\hat{L}_k^{T}\hat{L}_k^{*}\otimes I_{N}\Big).
\end{align}
From the theory of ordinary differential equations, the dynamical evolution is given by $|\rho(t)\rangle=e^{\hat{\mathcal{H}}t}|\rho(0)\rangle$.

\textbf{2. The decomposition of the non-Hermitian Hamiltonian.}

We decompose the $k$-local non-Hermitian Hamiltonian $\hat{\mathcal{H}}$ into a summation of unitaries with real coefficients such that
\begin{align}\label{NNHamiltonian}
\hat{\mathcal{H}}=\sum_{j=0}^{J-1}b_{j}\hat{\mathcal{H}}_{j},b_{j}>0,
\hat{\mathcal{H}}_{j}=\sigma_{a_{1}}^{(l_{1})}\cdots\sigma_{a_{i}}^{(l_{i})}\otimes\sigma_{a_{k}}^{(l_{k})},
\end{align}
where $\sigma_{a_{i}}^{(l_{i})}$ denotes the Pauli operator $\sigma_{a_{i}}$ acting on the $l_{i}$th qubit, $a_{i}\in\{x,y,z\}$. Note that in the tensor product of $\hat{\mathcal{H}}_{j}$ we have ignored the identity operators acting on the rest $2n-k$ qubits. Consider a general Hamiltonian which can be decomposed into a summation of Pauli strings
$\hat{\mathcal{H}}=\sum_{j=0}^{J_{1}-1}(c_{j}+\iota d_{j})\hat{\mathcal{P}}_{j}$ with complex coefficients, where $c_{j}\in\mathbb{R}$, $d_{j}\in\mathbb{R}$ and $\hat{\mathcal{P}}_{j}$ are unitaries. We have
\begin{align}
\hat{\mathcal{H}}&=\sum_{j=0}^{J_{1}-1}c_{j}\hat{\mathcal{P}}_{j}+\sum_{j=0}^{J_{1}-1}d_{j}(\iota\hat{\mathcal{P}}_{j})
=\sum_{j=0}^{J_{1}-1}|c_{j}|(-1)^{g(c_{j})}\hat{\mathcal{P}}_{j}
+\sum_{j=0}^{J_{1}-1}|d_{j}|[\iota(-1)^{g(d_{j})}\hat{\mathcal{P}}_{j}]
=\sum_{j=0}^{J-1}b_{j}\hat{\mathcal{H}}_{j},
\end{align}
where the function $g(x)=1$ $(0)$ if $x<0$ $(x>0)$,
\begin{equation}
b_{j}=\left\{\begin{aligned}
& |c_{j}|, & j=0,\cdots,J_{1}-1 \\
& |d_{j-J_{1}}|, & j=J_{1},\cdots,2J_{1}-1
\end{aligned}\right.,\hat{\mathcal{H}}_{j}=\left\{\begin{aligned}
& (-1)^{g(c_{j})}\hat{\mathcal{P}}_{j}, & j=0,\cdots,J_{1}-1 \\
& \iota(-1)^{g(d_{j-J_{1}})}\hat{\mathcal{P}}_{j-J_{1}}, & j=J_{1},\cdots,2J_{1}-1.
\end{aligned}\right.
\end{equation}
It is observed that $J=2J_{1}$ in Eq. (\ref{NNHamiltonian}).

\textbf{3. The decomposition of the block diagonal unitary.}

Decompose the block diagonal unitary $\Lambda_{\mathcal{Q}}$ into a product form,
\begin{align}
\Lambda_{\mathcal{Q}}=\sum_{j=0}^{J}|j\rangle\langle j|\otimes\mathcal{Q}_{j}=\prod_{j=0}^{J}\mathcal{C}_{m}^{|j\rangle}\mathcal{Q}_{j},
\end{align}
where the multi-qubits controlled unitary
\begin{align}
\mathcal{C}_{m}^{|j\rangle}\mathcal{Q}_{j}=|j\rangle\langle j|\otimes\mathcal{Q}_{j}
+\sum_{j\neq j^{'}=0}^{J}|j^{'}\rangle\langle j^{'}|\otimes I_{2^{k}},
\end{align}
$m$ denotes the number of controlled qubits and $|j\rangle$ in $\mathcal{C}_{m}^{|j\rangle}\mathcal{Q}_{j}$ represents the control basis, and $\mathcal{Q}_{j}$ is $k$-local for all $j$. It is convenient to write the state $|j\rangle$ by using the binary representation $j=j_{1}j_{2}\cdots j_{g}\cdots j_{m}$, where $j_{g}=0$ or $1$ for $g=1,2,\cdots,m$. Following the idea of the work \cite{Barenco1995Elementary}, we require to transform an arbitrary unitary $\mathcal{C}_{m}^{|j\rangle}\mathcal{Q}_{j}$ controlled by the state $|j\rangle=|j_{1}j_{2}\cdots j_{m}\rangle$ into a unitary
\begin{align}
\mathcal{C}_{m}^{|J\rangle}\mathcal{Q}_{j}
=|J\rangle\langle J|\otimes\mathcal{Q}_{j}+\sum_{j^{'}=0}^{J-1}|j^{'}\rangle\langle j^{'}|\otimes I_{2^{k}},~~|J\rangle=|11\cdots1\rangle,
\end{align}
controlled by the state $|J\rangle$. In particular, we have
\begin{align}
\mathcal{C}_{m}^{|j\rangle}\mathcal{Q}_{j}
=(\sigma_{x}^{!j_{1}}\otimes\sigma_{x}^{!j_{2}}\otimes\cdots\sigma_{x}^{!j_{m}}\otimes I_{2^{k}})\mathcal{C}_{m}^{|J\rangle}\mathcal{Q}_{j}
(\sigma_{x}^{!j_{1}}\otimes\sigma_{x}^{!j_{2}}\otimes\cdots\sigma_{x}^{!j_{m}}\otimes I_{2^{k}})^{\dag},
\end{align}
where $!j_{g}$ represents the NOT operator that returns $1$ $(0)$ when $j_{g}=0$ $(1)$, respectively. For instance, the matrix notation of the transformation $\mathcal{C}_{m}^{|j\rangle}\mathcal{Q}_{j}\rightarrow\mathcal{C}_{m}^{|J\rangle}\mathcal{Q}_{j}$ is
\begin{align}
&\mathcal{C}_{m}^{|j\rangle}\mathcal{Q}_{j}=
\begin{bmatrix}
I_{2^{k}} &  &  &  &  &  & \\
 & \ddots &  &  &  &  & \\
 &  & I_{2^{k}} &  &  &  & \\
 &  &  & \mathcal{Q}_{j} &  &  & \\
 &  &  &  & I_{2^{k}} &  & \\
 &  &  &  &  & \ddots & \\
 &  &  &  &  &  & I_{2^{k}}
\end{bmatrix}\mapsto
\begin{bmatrix}
I_{2^{k}} &  &  &  &  &  & \\
 & \ddots &  &  &  &  & \\
 &  & I_{2^{k}} &  &  &  & \\
 &  &  & I_{2^{k}} &  &  & \\
 &  &  &  & I_{2^{k}} &  & \\
 &  &  &  &  & \ddots & \\
 &  &  &  &  &  & \mathcal{Q}_{j}
\end{bmatrix}=\mathcal{C}_{m}^{|J\rangle}\mathcal{Q}_{j}.
\end{align}
Based on the decomposition Eq. (\ref{NNHamiltonian}), the $(m+k)$-qubit operator $\mathcal{C}_{m}^{|J\rangle}\mathcal{Q}_{j}$ can be expressed as $k$ $(m+2)$-qubit operators $\mathcal{C}_{m}^{|J\rangle}\sigma_{a_{1}},\cdots,
\mathcal{C}_{m}^{|J\rangle}\sigma_{a_{i}},\cdots,\mathcal{C}_{m}^{|J\rangle}\sigma_{a_{k}}$. In order to simulate the unitary $\mathcal{C}_{m}^{|J\rangle}\sigma_{a_{i}}$ with basic operators (single- and CNOT gates), we here review some preliminaries investigated in the work \cite{Barenco1995Elementary}.

\newtheorem{lemma}{Lemma}
\begin{lemma}\label{Lemma1}
For any $W\in\textrm{SU}(2)$ and a simulation error $\varepsilon>0$, the $m$-qubit controlled unitary $\mathcal{C}_{m}^{|J\rangle}W$ can be exactly simulated in terms of $\Theta(m^{2})$ basic operators. Furthermore, $\mathcal{C}_{m}^{|J\rangle}W$ can be approximately simulated within $\varepsilon$ by $\Theta(m\log_{2}(\varepsilon^{-1}))$ basic operators \cite{Barenco1995Elementary}.
\end{lemma}

Based on the Lemma 1, we obtain the following result.
\begin{lemma}\label{Lemma2}
For a $k$-local non-Hermitian operator $\mathcal{Q}(\Delta t)=\sum_{j=0}^{J}a_{j}\mathcal{Q}_{j}$, $J+1=2^{m}$, the block diagonal unitary $\Lambda_{\mathcal{Q}}=\sum_{j=0}^{J}|j\rangle\langle j|\otimes\mathcal{Q}_{j}$ can be exactly or approximately simulated with error $\epsilon_{\mathcal{Q}}$ via $\mathcal{O}(2^{m}km^{2})$ or $\mathcal{O}[2^{m}km\log_{2}(\epsilon_{\mathcal{Q}}^{-1})]$ basic operations.
\end{lemma}
\emph{Proof.} The unitary $\Lambda_{\mathcal{Q}}$ can be expressed as $J+1=2^{m}$ $m$-qubit controlled unitary $\mathcal{C}_{m}^{|J\rangle}\mathcal{Q}_{j}$. According to the locality of $\hat{\mathcal{H}}$ in Eq. (\ref{NNHamiltonian}), the non-unitary operator $\mathcal{Q}(\Delta t)$ is also $k$-local, which implies that the unitary $\mathcal{Q}_{j}$ acts on at most $k$-qubit. As a result, each unitary $\mathcal{C}_{m}^{|J\rangle}\mathcal{Q}_{j}$ consists of $k$ $(m+2)$-qubit operator $\mathcal{C}_{m}^{|J\rangle}\sigma_{a_{i}}^{(l_{i})}$. Based on the Lemma \ref{Lemma1}, the cost of approximate or exact simulating the unitary $\mathcal{C}_{m}^{|J\rangle}\sigma_{a_{i}}^{(l_{i})}$ within $\epsilon_{\mathcal{Q}}$ is $\Theta\left(m\log_{2}(\epsilon_{\mathcal{Q}}^{-1})\right)$ or $\Theta(m^{2})$. Thus, the total cost of simulating the unitary gate $\Lambda_{\mathcal{Q}}$ is $\mathcal{O}\left(2^{m}km\log_{2}(\epsilon_{\mathcal{Q}}^{-1})\right)$ or $\mathcal{O}(2^{m}km^{2})$.$\Box$

Here, we consider another decomposition of the unitary $\Lambda_{\mathcal{Q}}$, which is useful to compile $\Lambda_{\mathcal{Q}}$ with a low-depth unitary $V(\boldsymbol{\beta})$. Denote $j=j_1\cdots j_{g}\cdots j_{m}$. We have
\begin{align}
\Lambda_{\mathcal{Q}}&=\sum_{j=0}^{J}|j\rangle\langle j|\otimes\mathcal{Q}_{j}\nonumber\\
&=|j_{1}\rangle\langle j_{1}|\otimes\cdots\otimes|j_{g}\rangle\langle j_{g}|\otimes\cdots
\otimes|j_{m}\rangle\langle j_{m}|\otimes\mathcal{Q}_{j}\nonumber\\
&=\sum_{j=0}^{J}\frac{1}{2^{m}}\Big[(I_{2}+(-1)^{j_{1}}\sigma_{z})\otimes\cdots\otimes(I_{2}+(-1)^{j_{g}}\sigma_{z})
\otimes\cdots\otimes(I_{2}+(-1)^{j_{m}}\sigma_{z})\Big]\otimes\mathcal{Q}_{j}\nonumber\\
&=\sum_{j=0}^{J}\sum_{g=0}^{J}h_{g}^{j}\Lambda_{Q}(g),
\end{align}
where unitary $\Lambda_{Q}(g)$ consists of Pauli operator $\sigma_{z}$, the identity $I_2$, and $\mathcal{Q}_{j}$, and the coefficients $h_{g}^{j}$ is $2^{-m}$ or $-2^{-m}$. Thus we can estimate the cost function $G(\boldsymbol{\beta})=1-|\textrm{Tr}\left(\Lambda_{\mathcal{Q}}V^{\dag}
(\boldsymbol{\beta})\right)|^{2}/4^{2n+m}$ by calculating $\textrm{Tr}[\Lambda_{\mathcal{Q}}V^{\dag}(\boldsymbol{\beta})]$.

\textbf{4. The details on example 2.}

The Hamiltonian of the DTFIM on two sites is
$\mathcal{H}=-V\sigma_{z}^{(1)}\sigma_{z}^{(2)}-\Omega(\sigma_{x}^{(1)}\otimes I_{2}^{(2)}+I_{2}^{(1)}\otimes\sigma_{x}^{(2)})$ and the Lindblad operators $\hat{L}_{k}=\sqrt{\gamma}\sigma_{-}^{(k)}=\sqrt{\gamma}(\sigma_{x}^{(k)}-\iota\sigma_{y}^{(k)})$ for $k=1,2$. Thus, the master equation is
\begin{align}
\frac{d\rho(t)}{dt}=-\iota[\mathcal{H},\rho(t)]+\Bigg[&\hat{L}_{1}\rho(t)\hat{L}_{1}^{\dag}
-\frac{1}{2}\{\hat{L}_{1}^{\dag}\hat{L}_{1},\rho(t)\}+\hat{L}_{2}\rho(t)\hat{L}_{2}^{\dag}
-\frac{1}{2}\{\hat{L}_{2}^{\dag}\hat{L}_{2},\rho(t)\}\Bigg].
\end{align}
Using the Choi-Jamiolkowski isomorphism \cite{Havel2003Robust}, the master equation can be rewritten as a stochastic Schr$\ddot{\textrm{o}}$dinger equation form, $|\dot{\rho}(t)\rangle=\hat{\mathcal{H}}|\rho(t)\rangle$ on a $4$-qubit Hilbert space, where the non-Hermitian Hamiltonian
\begin{align}
\hat{\mathcal{H}}&=-\iota(I_{2}^{(1)}\otimes I_{2}^{(2)}\otimes\mathcal{H}-\mathcal{H}^{T}\otimes I_{2}^{(3)}\otimes I_{2}^{(4)})\nonumber\\
&+[\hat{L}_{1}^{*}\otimes I_{2}^{(2)}]\otimes[\hat{L}_{3}\otimes I_{2}^{(4)}]
-\frac{1}{2}[I_{2}^{(1)}\otimes I_{2}^{(2)}]\otimes[\hat{L}_{3}^{\dag}\hat{L}_{3}\otimes I_{2}^{(4)}]
-\frac{1}{2}[\hat{L}_{1}^{T}\hat{L}_{1}^{*}\otimes I_{2}^{(2)}]\otimes[I_{2}^{(3)}\otimes I_{2}^{(4)}]\nonumber\\
&+[I_{2}^{(1)}\otimes\hat{L}_{2}^{*}]\otimes[I_{2}^{(3)}\otimes\hat{L}_{4}]
-\frac{1}{2}[I_{2}^{(1)}\otimes I_{2}^{(2)}]\otimes[I_{2}^{(3)}\otimes\hat{L}_{4}^{\dag}\hat{L}_{4}]
-\frac{1}{2}[I_{2}^{(1)}\otimes\hat{L}_{2}^{T}\hat{L}_{2}^{*}]\otimes[I_{2}^{(3)}\otimes I_{2}^{(4)}]\nonumber\\
&=\iota[I_{2}^{(1)}\otimes I_{2}^{(2)}]\otimes[V\sigma_{z}^{(3)}\otimes\sigma_{z}^{(4)}+
\Omega(\sigma_{x}^{(3)}\otimes I_{2}^{(4)}+I_{2}^{(3)}\otimes\sigma_{x}^{(4)})]\nonumber\\
&-\iota[V\sigma_{z}^{(1)}\otimes\sigma_{z}^{(2)}+
\Omega(\sigma_{x}^{(1)}\otimes I_{2}^{(2)}+I_{2}^{(1)}\otimes\sigma_{x}^{(2)})]\otimes[I_{2}^{(3)}\otimes I_{2}^{(4)}]\nonumber\\
&+[\hat{L}_{1}^{*}\otimes I_{2}^{(2)}]\otimes[\hat{L}_{3}\otimes I_{2}^{(4)}]
-\frac{1}{2}[I_{2}^{(1)}\otimes I_{2}^{(2)}]\otimes[\hat{L}_{3}^{\dag}\hat{L}_{3}\otimes I_{2}^{(4)}]
-\frac{1}{2}[\hat{L}_{1}^{T}\hat{L}_{1}^{*}\otimes I_{2}^{(2)}]\otimes[I_{2}^{(3)}\otimes I_{2}^{(4)}]\nonumber\\
&+[I_{2}^{(1)}\otimes\hat{L}_{2}^{*}]\otimes[I_{2}^{(3)}\otimes\hat{L}_{4}]
-\frac{1}{2}[I_{2}^{(1)}\otimes I_{2}^{(2)}]\otimes[I_{2}^{(3)}\otimes\hat{L}_{4}^{\dag}\hat{L}_{4}]
-\frac{1}{2}[I_{2}^{(1)}\otimes\hat{L}_{2}^{T}\hat{L}_{2}^{*}]\otimes[I_{2}^{(3)}\otimes I_{2}^{(4)}]\nonumber,
\end{align}
with $\hat{L}_{3}=\sqrt{\gamma}\sigma_{-}^{(3)}$ and $\hat{L}_{4}=\sqrt{\gamma}\sigma_{-}^{(4)}$. Since $\hat{L}_{k}=\sqrt{\gamma}\sigma_{-}^{(k)}$, we have \begin{align}
\hat{L}_{k}^{*}=\hat{L}_{k}=\sqrt{\gamma}\sigma_{-}^{(k)}=\sqrt{\gamma}(\sigma_{x}^{(k)}-\iota\sigma_{y}^{(k)}),
\end{align}
and $\hat{L}_{k}^{T}=\hat{L}_{k}^{\dag}=\sqrt{\gamma}(\sigma_{x}^{(k)}+\iota\sigma_{y}^{(k)})$. Therefore, we obtain the following decompositions,
\begin{align}
\hat{L}_{1}^{*}\otimes\hat{L}_{3}&=\gamma(\sigma_{x}^{(1)}
\otimes\sigma_{x}^{(3)}-\iota\sigma_{x}^{(1)}\otimes\sigma_{y}^{(3)}
-\iota\sigma_{y}^{(1)}\otimes\sigma_{x}^{(3)}-\sigma_{y}^{(1)}\otimes\sigma_{y}^{(3)}),\nonumber\\
-\frac{1}{2}\hat{L}_{3}^{\dag}\hat{L}_{3}&=\gamma(-I_{2}^{(3)}-\sigma_{z}^{(3)}),~~
-\frac{1}{2}\hat{L}_{1}^{T}\hat{L}_{1}^{*}=\gamma(-I_{2}^{(1)}-\sigma_{z}^{(1)}),\nonumber\\
\hat{L}_{2}^{*}\otimes\hat{L}_{4}&=\gamma(\sigma_{x}^{(2)}\otimes\sigma_{x}^{(4)}-\iota\sigma_{x}^{(2)}\otimes\sigma_{y}^{(4)}
-\iota\sigma_{y}^{(2)}\otimes\sigma_{x}^{(4)}-\sigma_{y}^{(2)}\otimes\sigma_{y}^{(4)}),\nonumber\\
-\frac{1}{2}\hat{L}_{4}^{\dag}\hat{L}_{4}&=\gamma(-I_{2}^{(4)}-\sigma_{z}^{(4)}),~~
-\frac{1}{2}\hat{L}_{2}^{T}\hat{L}_{2}^{*}=\gamma(-I_{2}^{(2)}-\sigma_{z}^{(2)}).\nonumber
\end{align}
It is clear that $\hat{\mathcal{H}}$ can be expressed as a summarization of $19$ unitary operators (see Table S1), and the coefficients $b_j$ are given by
\begin{align}
b=(b_{0},b_{1},\cdots,b_{18})=(V,\Omega,\Omega,V,\Omega,\Omega,\gamma,\gamma,\gamma,\gamma,\gamma,\gamma,\gamma,\gamma,\gamma,\gamma,\gamma,\gamma,4\gamma).
\end{align}

\begin{table}[h!]
\begin{center}
\label{Example2LCU}
\begin{tabular}{|c|c|c|c|c|c|}
\hline
$k$ & 0 & 1 & 2 & 3 & 4\\
\hline
$\hat{\mathcal{H}}_{k}$ & $\iota\sigma_{z}^{(3)}\otimes\sigma_{z}^{(4)}$ & $\iota\sigma_{x}^{(3)}$ & $\iota\sigma_{x}^{(4)}$
& $-\iota\sigma_{z}^{(1)}\otimes\sigma_{z}^{(2)}$ & $-\iota\sigma_{x}^{(1)}$ \\
\hline
$k$ & 5 & 6 & 7 & 8 & 9 \\
\hline
$\hat{\mathcal{H}}_{k}$ & $-\iota\sigma_{x}^{(2)}$ & $\sigma_{x}^{(1)}\otimes\sigma_{x}^{(3)}$ & $-\iota\sigma_{x}^{(1)}\otimes\sigma_{y}^{(3)}$
& $-\iota\sigma_{y}^{(1)}\otimes\sigma_{x}^{(3)}$ & $-\sigma_{y}^{(1)}\otimes\sigma_{y}^{(3)}$ \\
\hline
$k$ & 10 & 11 & 12 & 13 & 14 \\
\hline
$\hat{\mathcal{H}}_{k}$ & $-\sigma_{z}^{(3)}$ & $-\sigma_{z}^{(1)}$ & $\sigma_{x}^{(2)}\otimes\sigma_{x}^{(4)}$
& $-\iota\sigma_{x}^{(2)}\otimes\sigma_{y}^{(4)}$ & $-\iota\sigma_{y}^{(2)}\otimes\sigma_{x}^{(4)}$ \\
\hline
$k$ & 15 & 16 & 17 & 18 &  \\
\hline
$\hat{\mathcal{H}}_{k}$ & $-\sigma_{y}^{(2)}\otimes\sigma_{y}^{(4)}$ & $-\sigma_{z}^{(4)}$ & $-\sigma_{z}^{(2)}$
& $-I_{16}$ &  \\
\hline
\end{tabular}
\end{center}
\caption{The unitary operators of the non-Hermitian Hamiltonian $\hat{\mathcal{H}}$ in the LCU decomposition.}
\end{table}

The first order approximation of the exponential operator $\mathcal{U}(\Delta t)=e^{\hat{\mathcal{H}}\Delta t}$ is
\begin{align}
\mathcal{Q}(\Delta t)=I_{16}+\hat{\mathcal{H}}\Delta t=I_{16}+\sum_{j=0}^{18}\Delta tb_{j}\hat{\mathcal{H}}_{j}.
\end{align}
Since the number of unitary operators is $19$ which is not a power of $2$, we require to divide the identity $I_{16}$ into $13$ sub-terms such as
\begin{align}
\mathcal{Q}(\Delta t)=\sum_{j=0}^{18}\Delta tb_{j}\hat{\mathcal{H}}_{j}+\underbrace{\frac{1}{13}I_{16}+\cdots+\frac{1}{13}I_{16}}_{13}.
\end{align}
We now have reexpression, $\mathcal{Q}(\Delta t)=\sum_{j=0}^{31}a_{j}\mathcal{Q}_{j}$, where the coefficients $a_{j}>0$ and unitary operators $\mathcal{Q}_{j}$ are given by
\begin{equation}
a_{j}=\left\{\begin{aligned}
& \Delta tb_{j}, & j=0,1,\cdots,18, \\
& \frac{1}{13}, & j=19,\cdots,31,
\end{aligned}\right.
~~\textrm{and}~~\mathcal{Q}_{j}=\left\{\begin{aligned}
& \hat{\mathcal{H}}_{j}, & j=0,1,\cdots,18, \\
& I_{16}, & j=19,\cdots,31.
\end{aligned}\right.
\end{equation}
The number of unitaries is $32=2^{5}$. Thus in our numerical simulation, the ancillary system requires $5$ qubits to store the superposition state $|a\rangle=\sum_{j=0}^{31}\sqrt{\frac{a_{j}}{A}}|j\rangle$, where $A=\sum_{j=0}^{31}a_{j}$.

\textbf{5. The simulation of a general quantum channel.}

We here report that the simulation approach can be naturally generalized to simulate a general quantum channel. The Kraus representation of a quantum channel can be written as \cite{Nielsen2000Quantum},
\begin{align}\label{Channel}
\rho\mapsto\mathcal{E}(\rho)=\sum_{l=1}^{L}E_{l}\rho E_{l}^{\dag},~~\rho\in\mathbb{C}^{N\times N},~~N=2^{n},
\end{align}
with $1\leq L\leq N^{2}$ and $\sum_{l=1}^{L}E_{l}^{\dag}E_{l}=I_{N}$, where $I_{N}$ is the identity matrix. The master equation in the Lindblad form can be expressed in terms of the quantum channel formalism,
\begin{align}
\dot{\rho}(t)&=-\iota[\mathcal{H},\rho(t)]+\sum_{k}[\hat{L}_{k}\rho(t)\hat{L}_{k}^{\dag}
-\frac{1}{2}\{\hat{L}_{k}^{\dag}\hat{L}_{k},\rho(t)\}]\nonumber\\
&=-\iota\Big[\mathcal{H}\rho(t)-\rho(t)\mathcal{H}\Big]
+\sum_{k}\Big[\hat{L}_{k}\rho(t)\hat{L}_{k}^{\dag}
-\frac{1}{2}\hat{L}_{k}^{\dag}\hat{L}_{k}\rho(t)
-\frac{1}{2}\rho(t)\hat{L}_{k}^{\dag}\hat{L}_{k}\Big].
\end{align}
When $\Delta t\rightarrow0$, the master equation becomes
\begin{align}
\dot{\rho}(t)&=\frac{\rho(t+\Delta t)-\rho(t)}{\Delta t}
=-\iota\Big[\mathcal{H}\rho(t)-\rho(t)\mathcal{H}\Big]
+\sum_{k}\Big[\hat{L}_{k}\rho(t)\hat{L}_{k}^{\dag}
-\frac{1}{2}\hat{L}_{k}^{\dag}\hat{L}_{k}\rho(t)
-\frac{1}{2}\rho(t)\hat{L}_{k}^{\dag}\hat{L}_{k}\Big].
\end{align}
Consider the non-unitary channels given by the Kraus operators $E_{0}$ and $E_{k}$,
\begin{align}
\mathcal{E}_{0}[\rho(t)]&=E_{0}\rho(t)E_{0}^{\dag}
=\Big(I_{N}-\iota\mathcal{H}\Delta t-\frac{\Delta t}{2}\sum_{k}\hat{L}_{k}^{\dag}\hat{L}_{k}\Big)\rho(t)
\Big(I_{N}+\iota\mathcal{H}\Delta t-\frac{\Delta t}{2}\sum_{k}\hat{L}_{k}^{\dag}\hat{L}_{k}\Big)\nonumber\\
&=\rho(t)-\iota\Delta t[\mathcal{H},\rho(t)]-\Delta t\sum_{k}\Big[\frac{1}{2}\hat{L}_{k}^{\dag}\hat{L}_{k}\rho(t)
+\frac{1}{2}\rho(t)\hat{L}_{k}^{\dag}\hat{L}_{k}\Big]
\end{align}
and
\begin{align}
\mathcal{E}_{k}[\rho(t)]&=E_{k}\rho(t)E_{k}^{\dag}
=\Big(\sqrt{\Delta t}\hat{L}_{k}\Big)\rho(t)\Big(\sqrt{\Delta t}\hat{L}_{k}\Big)^{\dag}=\Delta t\hat{L}_{k}\rho(t)\hat{L}_{k}^{\dag}.
\end{align}
We have the quantum channel formalism of the master equation, $\rho(t+\Delta t)=\mathcal{E}_{0}[\rho(t)]+\mathcal{E}_{k}[\rho(t)]$. Here the normalization condition for the Kraus operators holds for an infinitesimal time $\Delta t$,
\begin{align}
E_{0}^{\dag}E_{0}+\sum_kE_k^{\dag}E_k&=\Big(I_{N}-\iota\mathcal{H}\Delta t-\frac{\Delta t}{2}\sum_{k}\hat{L}_{k}^{\dag}\hat{L}_{k}\Big)^{\dag}
\Big(I_{N}-\iota\mathcal{H}\Delta t-\frac{\Delta t}{2}\sum_{k}\hat{L}_{k}^{\dag}\hat{L}_{k}\Big)
+\sum_{k}\Big(\sqrt{\Delta t}\hat{L}_{k}\Big)^{\dag}\Big(\sqrt{\Delta t}\hat{L}_{k}\Big)\nonumber\\
&=I_{N}+\mathcal{O}(\Delta t^2).
\end{align}

In order to simulate the quantum channel $\mathcal{E}$ (\ref{Channel}), we consider the vectorized form
\begin{align}
|\rho\rangle=\sum_{l=1}^{L}E_{l}^{*}\otimes E_{l}|\rho\rangle.
\end{align}
By representing the operator $\sum_{l=1}^{L}E_{l}^{*}\otimes E_{l}$ in terms of unitaries, similar to the simulation of OQS, we can simulate the quantum channel in composite systems.
\end{document}